\documentclass[pre,twocolumn,amssymb]{revtex4-1}

\usepackage{mathtools}
\usepackage{amsmath}
\usepackage{float}
\usepackage{graphicx}
\usepackage{color}
\usepackage{epsfig}
\usepackage{latexsym}
\usepackage{bm}
\usepackage[english]{babel}
\usepackage[colorlinks=true, allcolors=blue]{hyperref}
\usepackage{color}
\usepackage[normalem]{ulem}

\begin{document}

\newcommand{\rar}{\rightarrow}
\newcommand{\lar}{\leftarrow}
\newcommand{\rlh}{\rightleftharpoons}
\newcommand{\eref}[1]{Eq.~(\ref{#1})}%
\newcommand{\Eref}[1]{Equation~(\ref{#1})}%
\newcommand{\fref}[1]{Fig.~\ref{#1}} %
\newcommand{\Fref}[1]{Figure~\ref{#1}}%
\newcommand{\sref}[1]{Sec.~\ref{#1}}%
\newcommand{\Sref}[1]{Section~\ref{#1}}%
\newcommand{\aref}[1]{Appendix~\ref{#1}}%

\renewcommand{\ni}{{\noindent}}
\newcommand{\dprime}{{\prime\prime}}
\newcommand{\be}{\begin{equation}}
\newcommand{\ee}{\end{equation}}
\newcommand{\bea}{\begin{eqnarray}}
\newcommand{\eea}{\end{eqnarray}}
\newcommand{\nn}{\nonumber}
\newcommand{\bk}{{\bf k}}
\newcommand{\bQ}{{\bf Q}}
\newcommand{\q}{{\bf q}}
\newcommand{\s}{{\bf s}}
\newcommand{\bN}{{\bf \nabla}}
\newcommand{\bA}{{\bf A}}
\newcommand{\bE}{{\bf E}}
\newcommand{\bj}{{\bf j}}
\newcommand{\bJ}{{\bf J}}
\newcommand{\bs}{{\bf v}_s}
\newcommand{\bn}{{\bf v}_n}
\newcommand{\bv}{{\bf v}}
\newcommand{\la}{\left\langle}
\newcommand{\ra}{\right\rangle}
\newcommand{\dg}{\dagger}
\newcommand{\br}{{\bf{r}}}
\newcommand{\brp}{{\bf{r}^\prime}}
\newcommand{\bq}{{\bf{q}}}
\newcommand{\hx}{\hat{\bf x}}
\newcommand{\hy}{\hat{\bf y}}
\newcommand{\bS}{{\bf S}}
\newcommand{\cU}{{\cal U}}
\newcommand{\cD}{{\cal D}}
\newcommand{\bR}{{\bf R}}
\newcommand{\pll}{\parallel}
\newcommand{\sumr}{\sum_{\vr}}
\newcommand{\cP}{{\cal P}}
\newcommand{\cQ}{{\cal Q}}
\newcommand{\cS}{{\cal S}}
\newcommand{\ua}{\uparrow}
\newcommand{\da}{\downarrow}
\newcommand{\red}{\textcolor {red}}
\newcommand{\blue}{\textcolor {blue}}
\newcommand{\1}{{\oldstylenums{1}}}
\newcommand{\2}{{\oldstylenums{2}}}
\newcommand{\mDelta}{\varepsilon}
\newcommand{\m}{\tilde m}
\def\lsim {\protect \raisebox{-0.75ex}[-1.5ex]{$\;\stackrel{<}{\sim}\;$}}
\def\gsim {\protect \raisebox{-0.75ex}[-1.5ex]{$\;\stackrel{>}{\sim}\;$}}
\def\lsimeq {\protect \raisebox{-0.75ex}[-1.5ex]{$\;\stackrel{<}{\simeq}\;$}}
\def\gsimeq {\protect \raisebox{-0.75ex}[-1.5ex]{$\;\stackrel{>}{\simeq}\;$}}

\title{\large{Transport and fluctuations in mass aggregation processes: mobility-driven clustering}}

\author{Subhadip Chakraborti$^{1,2}$} 
\email{subhadip.chakraborti@icts.res.in}
\author{Tanmoy Chakraborty$^{1}$}
\email{tanmoy.chakraborty@bose.res.in}  
\author{Arghya Das$^{2}$}
\email{arghya.das@icts.res.in}
\author{Rahul Dandekar$^{3,4}$}
\author{Punyabrata Pradhan$^1$}

\affiliation{$^1$Department of Theoretical Sciences, S. N. Bose National Centre for Basic Sciences, Block-JD, Sector-III, Salt Lake, Kolkata 700106, India \\ $^2$International Centre for Theoretical Sciences, Tata Institute of Fundamental Research, Bengaluru 560089, India \\ $^3$The Institute of Mathematical Sciences, C.I.T. Campus, Taramani, Chennai 600113, India \\ $^4$Homi Bhabha National Institute, Training School Complex, Anushakti Nagar, Mumbai 400094, India }

\begin{abstract}

\noindent{ We calculate the bulk-diffusion coefficient and the conductivity in nonequilibrium conserved-mass aggregation processes on a ring.  These processes involve chipping and fragmentation of masses, which diffuse on a lattice and aggregate with their neighboring masses upon contact, and, under certain conditions, they exhibit a condensation transition. We find that, even in the absence of microscopic time reversibility, the systems satisfy an Einstein relation, which connects the ratio of the conductivity and the bulk-diffusion coefficient to mass fluctuation. Interestingly, when aggregation dominates over chipping, the conductivity or, equivalently, the mobility of masses, is greatly enhanced.  The enhancement in the conductivity, in accordance with the Einstein relation, results in large mass fluctuations and can induce a mobility-driven clustering in the systems. 
Indeed, in a certain parameter regime, we show that the conductivity, along with the mass fluctuation, diverges beyond a critical density, thus characterizing the previously observed nonequilibrium condensation transition [Phys. Rev. Lett. {\bf 81}, 3691 (1998)] in terms of an instability in the conductivity.
Notably, the bulk-diffusion coefficient remains finite in all cases.  We find our analytic results in quite good agreement with simulations.  }
\typeout{polish abstract}
\end{abstract}

\maketitle

\section{Introduction}

Mass aggregation processes involving fragmentation, diffusion and aggregation
are ubiquitous in nature. They occur in a variety of growth and aggregation related phenomena, such as droplet and cloud formation \cite{droplets-1992,smoke}, planet and island formation  \cite{planets, river}, aggregation in colloidal suspensions \cite{colloids}, traffic flow \cite{traffic, Krug}, polymer gel and aerosol formation \cite{smoke, Ziff_PRL1982, Family_PRL1983}, self-assembly in nanomaterials \cite{VVV},  etc. These systems are inherently driven out of thermal equilibrium as they violate detailed balance due to the lack of time reversibility at the microscopic levels. 
Not surprisingly, for such systems, there is no unified statistical mechanics framework based on a general thermodynamic principle.

Throughout the past decades, significant efforts have been made to understand various static and dynamic properties of aggregation processes through studies of simple models, which are easy to simulate on computers and amenable to analytical calculations. Unlike their equilibrium counterparts, these nonequilibrium model-systems, while having simple dynamical rules, possess nontrivial spatio-temporal structures. Indeed, under certain conditions, they exhibit striking collective behaviors, such as cluster and pattern formation \cite{droplets-1992}, giant mass fluctuations and intermittency \cite{Barma_PRL2013, Barma_JSP2014}, gelation \cite{Ziff_PRL1982} and condensation transition \cite{Krapivsky_PRE1996, Barma_PRL1998, Barma_JSP2000}, etc. 
In this paper, we aim to characterize some of the above mentioned collective properties in terms of the two transport coefficients - the bulk-diffusion coefficient and the conductivity.

In some of the earliest studies, the clustering properties were explored through simple kinetic models of aggregation related growth processes, such as polymerization \cite{Ziff_PRL1982, Leyvraz_JPhysA1982, Family_PRL1983} and droplet formation \cite{Family-Meakin_PRL1988}, etc. Later, several variants of these models were introduced through generalized fragmentation and aggregation  kernels, which specify the rates with which masses get fragmented and aggregate \cite{Ziff_PRL1982, Ziff_PRB1988, Krapivsky_PRE1996}. In fact, depending on the relative strength of fragmentation and aggregation processes, the systems undergo gelation or condensation transition and exhibit self-similarity, which are manifested in the power-law cluster size distributions of masses. 
Although, in a natural environment, these mass aggregation processes can be dominated by diffusion \cite{Barkai_pre2020}, the earlier studies however did not take into account the underlying spatial structures through which the mass diffusion could occur in these systems.

Indeed, in a more realistic setting, the process of diffusion should be considered to fully characterize the spatio-temporal properties of mass aggregation processes. To this end, the diffusion was incorporated in Refs. \cite{Barma_PRL1998, Barma_JSP2000, Rajesh_PRE2002}, where masses, in addition to being fragmented with a certain rate, can also diffuse around and aggregate when a mass comes into contact with any other neighboring diffusing mass. Interestingly, in this case, the system was shown to exhibit, beyond a critical global density, a condensation transition and diverging mass fluctuations. However, in these works, the fragmentation was considered only through chipping of a single-unit mass.  Later, a generalized version of the fragmentation processes, though without any diffusion, was considered in models where arbitrary amounts of mass can get fragmented \cite{Puri}; however, in the absence of diffusion and aggregation, the latter models do not exhibit any condensation transition.

While such mass aggregation processes provide  a simple but a novel mechanism of a nonequilibrium condensation transition, the dynamical origin of the phase transition, i.e., exactly how mass transport affects the mass fluctuations, especially near the transition point, is not well understood.
More generally, hydrodynamics of these mass aggregation processes and the related time-dependent properties, such as that of density relaxations in the systems, are still largely unexplored, even as they are of a significant interest because hydrodynamics can also characterize large-scale fluctuations in these systems \cite{Bertini_PRL2001, Bertini_RMP2015}. 
Deriving hydrodynamics of driven many-body systems is of fundamental interest in statistical physics and requires calculations of density-dependent transport coefficients, such as the bulk-diffusion coefficient and the conductivity. However, the problem in general remains a challenging one \cite{Landim, Spohn}.  The difficulties arise due to mainly two reasons. Firstly, interacting particle systems can have a ``non-gradient'' structure \cite{Mallick_PRE2014}, making it hard to find a coarse-grained local current, which can be expressed as a gradient of a local observable. Secondly, unlike in equilibrium, the steady-state probability weights of microscopic configurations in most cases are a-priori not known  and calculating the average of the local observables, required to obtain the transport coefficients,  is not particularly easy.

In this paper, we derive hydrodynamics of a broad class of nonequilibrium conserved-mass aggregation processes on a ring of discrete lattice sites and explore the relationship between fluctuations and transport in the systems. To this end, we consider a generalized version of the aggregation models, where all three processes -  fragmentation, diffusion and aggregation of masses - are present. In addition to the chipping of a single-unit mass, we introduce fragmentation processes, where a random amount of mass can get detached from the parent mass. The fragmented mass diffuses symmetrically, to their right or left with equal probability, and aggregates upon contact with a neighboring mass, if there is any. The total mass remains conserved in the system. For simplicity, we consider fragmentation, diffusion and aggregation rates being  independent of masses at departure (or destination) sites.

The hydrodynamic time evolution of  the local density field in these systems is governed by the two transport coefficients - the bulk-diffusion coefficient and the conductivity. In a few special cases, including the most interesting  one exhibiting a condensation transition, the transport coefficients are analytically calculated from the knowledge of the single-site mass distributions; in general, the transport coefficients in these mass aggregation processes are non-linear functions of density. Indeed, the calculations of the transport coefficients have been made possible due to the following simplifying features of the models considered here. The systems satisfy a ``gradient'' property, implying that the instantaneous coarse-grained local current can be expressed as a gradient of a local observable. Moreover, we find that the systems have, in the limit of large system sizes, vanishingly small spatial correlations so that we can use a mean-field theory to exactly calculate the local observables required to obtain the transport coefficients.
However, for the generic parameter values, we have calculated the transport coefficients only numerically. Remarkably, despite the violation of detailed balance, the ratio of the conductivity to the bulk-diffusion coefficient is related to the mass fluctuation through an equilibrium-like  Einstein relation [see Eq.~(\ref{ER})].

 Notably fragmentation, diffusion and aggregation processes together greatly enhance the conductivity, or equivalently the mobility of masses, resulting in large mass fluctuations and a {\it mobility-driven clustering} in the systems. 
Indeed, in a certain parameter regime, where fragmentation and aggregation dominate over single-particle chipping, the systems undergo  a dynamic phase transition in the sense that the conductivity diverges in this regime (or, equivalently, the resistivity vanishes). 
To characterize the collective dynamical behaviors of such systems, we parameterize the fragmentation processes through a probability distribution $\phi(v)$, which is defined over non-negative integers $v \geq 0$. During a fragmentation event, a random  $v$ units of mass get fragmented from a site, provided the mass at a site is greater than or equal to $v$, and the fragmented $v$ units of mass then diffuse together to one of the neighboring sites; if the mass at the site is less than $v$, the whole mass diffuses to the neighboring site.  Though the system can have a large number of parameters depending on the distribution function $\phi(v)$, we broadly observe two kinds of dynamical behaviors. If the typical value of $v$ is finite, there is {\it no} phase transition in the systems and the two transport coefficients - the bulk-diffusion coefficient and the conductivity - remain finite at all densities; however, the conductivity can be quite large if the typical value of $v$ is large. On the other hand, when the typical value of $v$ diverges, the systems undergo, beyond a critical density, a condensation transition, where a macroscopic-size mass condensate forms in the systems and subsystem mass fluctuations diverge. Dynamically, the condensation transition is characterized by the singularity in the conductivity, which also diverges at the transition point.

To quantify the transport and the fluctuation characteristics of the mass-aggregating systems in concrete terms, we consider two one-parameter families of probability distributions - a localized distribution $\phi(v) = \delta_{v,v_0}$ and an exponential distribution $\phi(v) \propto \exp(-v/v_*)$, where we vary the typical values $v_0$ and $v_*$ for any transition to occur. We find that, in the presence of chipping, the system exhibits a condensation transition  at a finite critical density $\rho_c$ only when $v_0 \rightarrow \infty$ (or $v_* \rightarrow \infty$); in the absence of chipping, a macroscopic condensate however forms at any nonzero density when $v_0, v_* \rightarrow \infty$. We show that, at the phase transition point, both the mass fluctuation and the conductivity develop a simple-pole singularity, i.e.,  both the quantities as a function of density $\rho$ diverge as $(\rho_c - \rho)^{-1}$.
Indeed the intimate connection between the transport and the fluctuation is precisely encoded in the Einstein relation. Interestingly, the bulk-diffusion coefficients in all cases remain bounded.

The paper is organized as follows. In Sec. \ref{Model}, we define the model where chipping of single unit of mass and fragmentation of a variable amount of mass $v$ are both allowed with certain rates. In Sec. \ref{general_hydrodynamics}, we describe the linear response theory for calculating the transport coefficients. In Sections \ref{Sec_VarI} and \ref{Sec_Var2}, we study the two types of fragmentation rules where the distribution  is either localized $\phi(v)=\delta_{v,v_0}$ or exponential $\phi(v) \propto \exp(-v/v_*)$, having a typical size $v_0$ or $v_*$, respectively. 
We conclude in Sec. \ref{conclusion}.

\section{Mass Aggregation Models}
\label{Model}

\noindent
In this section, we define a generalized version of mass aggregation processes, which have been studied intensively in the past \cite{Barma_PRL1998, Barma_JSP2000, Puri}.
The system consists of $L$ sites on a one dimensional ring, where a site $i$ is associated with a mass, or particle number, $m_i \in [0, 1, 2, \dots]$, taking unbounded integer values (mass at any site is measured in the unit of individual particle mass). In these processes, total mass $M=\sum_{i=1}^L m_i$ remains conserved with global density $\rho=M/L$ fixed. The dynamics evolves in a continuous time and, at any instant of time, there are two kinds of dynamical updates possible at an occupied site: 
\\
\\
(A) chipping of a single-unit mass with rate $p$, and
\\
\\
(B) fragmentation of  a random amount of mass with rate $q$. 
\\
\\
In the event of (A), a single-unit of mass, or a particle, is chipped off from the departure site, provided the site is occupied. The chipped-off mass is then transferred, symmetrically, to one of its nearest neighbors with equal probability $1/2$. In the event of (B), a random number $v$ is drawn from a probability distribution $\phi(v)$. Provided that the site has mass greater than $v$, the whole block of $v$ units of mass are fragmented and transferred together to one of the nearest neighbors with probability $1/2$; otherwise, the whole block of mass is transferred in the similar way, keeping the departure site empty. In  Monte Carlo simulations, we employ random sequential updates where a site is updated with a unit rate.
As shown later, large-scale behaviors of the system is determined by the competition between chipping and fragmentation, leading to a dynamical phase transition upon tuning either the density or the relative strength $\tilde q = q/(p+q)$ of fragmentation to chipping.

Let $m_i(t)$ be the mass at site $i$ and at time $t$. Now let us define the indicator function,
\begin{equation}\label{indicator-activity}
\hat{a}_i= 1-\delta_{m_i,0},
\end{equation}
which is $1$ if $i^{th}$ site is occupied and $0$ otherwise. We define another indicator function $\hat{a}_i^v$, which is $1$ if the $i^{th}$ site contains at least $v$ particles, and $0$ otherwise, 
\begin{eqnarray} 
\hat{a}_i^v =
\left\{
\begin{array}{ll}
 1~~ \mbox{if}~~ m_i \geq v, \\ 
 0~~ \mbox{if}~~ m_i < v.
\end{array}
\right.
\end{eqnarray} Then the continuous-time evolution for mass $m_i(t)$, in an infinitesimal time interval $dt$, can be written as given below,
\begin{eqnarray} \nonumber
m_i(t+dt) = ~~~~~~~~~~~~~~~~~~~~~~~~~~~~~~~~~~~~~~~~~~~~~~~~~~ \\ 
\left\{
\begin{array}{ll}
m_i(t) - 1            & {\rm prob.}~ p\hat{a}_i dt, \\
m_i(t) + 1            & {\rm prob.}~ p\hat{a}_{i-1} dt/2, \\
m_i(t) + 1            & {\rm prob.}~ p\hat{a}_{i+1} dt/2, \\
m_i(t) - v            & {\rm prob.}~ q\hat{a}_i^v \phi(v) dt, \\
0                     & {\rm prob.}~ q\hat{a}_i(1-\hat{a}_i^v) \phi(v) dt, \\
m_i(t) + m_{i-1}(t)   & {\rm prob.}~ q\hat{a}_{i-1}(1-\hat{a}_{i-1}^v) \phi(v) dt/2, \\
m_i(t) + v            & {\rm prob.}~ q\hat{a}_{i-1}^v \phi(v) dt/2, \\
m_i(t) + m_{i+1}(t)   & {\rm prob.}~ q\hat{a}_{i+1}(1-\hat{a}_{i+1}^v) \phi(v) dt/2, \\
m_i(t) + v            & {\rm prob.}~ q\hat{a}_{i+1}^v \phi(v) dt/2, \\
m_i(t)                & {\rm prob.}~ 1-\Sigma dt,
\end{array}
\right.
\label{General_Update}
\end{eqnarray}
where the sum of the rates for all possible mass-transfer events in the infinitesimal time interval $dt$ is given by
\begin{eqnarray}\label{sigma_update}
\Sigma =& \nonumber \frac{p}{2} (2\hat{a}_i + \hat{a}_{i-1}+ \hat{a}_{i+1}) + \frac{q}{2} \sum_{v=0}^\infty\phi(v) \left[ 2\hat{a}_i^v + 2\hat{a}_i(1-\hat{a}_i^v) \right. \\ 
&+ \left. \hat{a}_{i-1}(1-\hat{a}_{i-1}^v) + \hat{a}_{i-1}^v + \hat{a}_{i+1}(1-\hat{a}_{i+1}^v)+ \hat{a}_{i+1}^v \right]. ~~~~
\end{eqnarray}

\section{ Hydrodynamics: theoretical framework }
\label{general_hydrodynamics}

According to the mass aggregation model defined in Eq.~(\ref{General_Update}), there is only one conserved quantity, viz. the mass or the particle-numbers. Consequently, the hydrodynamic evolution of local density $\rho(x,\tau)$ at suitably scaled space and time coordinates $x$ and $\tau$, respectively, is governed by the two transport coefficients - the bulk-diffusion coefficient $D(\rho)$ and the particle conductivity $\chi(\rho)$, which are in general non-linear functions of density $\rho$. In this section, we set up a general framework to calculate the two transport coefficients directly from the microscopic dynamics. The bulk-diffusion coefficient can be calculated in an appropriate scaling limit from the time-evolution equation of the local density field obtained using the continuous-time microscopic dynamics as Eq.~(\ref{General_Update}). On the other hand, the conductivity can be obtained by calculating the response of the system against an external perturbation (force here), which couples to the mass of the particles. As in a standard linear response theory, we calculate the average current due to a small externally applied biasing force field, which drives the particles in a preferred direction. Indeed the conductivity calculated in this paper is analogous to the conductivity of charged particles in the presence of an electric field. We incorporate the effect of the biasing force into the microscopic dynamics [Eq.~(\ref{General_Update})] of the model by following a local detailed balance condition. In the presence of a biasing force of magnitude $F$, which is applied, say, in the counter-clockwise direction, the particle hopping rates are modified by exponentially weighting the original hopping rates of Eq.~(\ref{General_Update}) as given below \citep{Bertini_RMP2015},
\begin{eqnarray}
c^{F}_{i \rightarrow j} &=& c_{i \rightarrow j} \exp\left[ \frac{1}{2} \Delta m_{i \rightarrow j} F(j-i) \delta x \right],  \nonumber \\
&\simeq& c_{i \rightarrow j} \left[ 1 + \frac{1}{2} \Delta m_{i \rightarrow j} F(j-i) \delta x \right] +{\cal O}(F^2).
\label{bias4}
\end{eqnarray}
Here $c_{i \rightarrow j}$ and  $c^{F}_{i \rightarrow j}$  are the mass transfer rates from site $i$ to site $j=i \pm 1$ in the absence and in the presence of the biasing force of magnitude $F$, respectively, and $\Delta m_{i \rightarrow j}$ is the transferred mass from site $i$ to $j$, and $\delta x$ is the lattice spacing. 
Also, in the last step of the above equation, we have kept only the leading order term in $F$, which is required for the linear response analysis performed below.

Large-scale spatio-temporal properties of a system can be understood in terms of the relevant local degrees of freedom, which vary slowly in space and time; in the case of mass aggregation processes considered here, the desired slow variable is the local mass density 
\be 
\rho_i(t) = \langle m_i(t) \rangle.
\ee 
Interestingly, the large-scale fluctuation properties of a diffusive system can be characterized through the large-deviation probabilities for the coarse-grained local density and local diffusive and drift currents, which are obtained on a suitable macroscopic scale (i.e., the diffusive scaling limit discussed later) through a continuity equation corresponding to the conserved local density. This is the essence of a recently developed fluctuating hydrodynamics, or the macroscopic fluctuation theory, which provides a general framework for studying macroscopic fluctuations in the diffusive systems \cite{Bertini_JSP2002, Bertini_RMP2015}. In the past, for systems satisfying a ``gradient condition'' \cite{Mallick_PRE2014}, this particular approach has been elucidated for various systems that possess a {\it local equilibrium} property on large spatio-temporal scales \cite{Eyink-Lebowitz-Spohn1991, Bertini_PRL2001}. Later, it has been used to derive hydrodynamics for various conserved-mass transport processes that manifestly violate detailed balance at the microscopic level \cite{Das_PRE2017, Sayani_FES}. In a more recent development, large-scale hydrodynamics has been derived for systems having a ``generalized gradient property'' \cite{Krapivsky_JSTAT2018, Tanmoy_longhop}. 
The mass aggregation models considered here have the ``gradient property'', which, along with the hypothesis of a {\it local steady state} - analogous to that of local equilibrium \cite{Eyink-Lebowitz-Spohn1991, Bertini_PRL2001}, can be used to calculate the transport coefficients.

To this end, assuming the existence of local steady state, we introduce local single-site mass distribution ${\rm Prob.}(m_i=m|\rho_i) \equiv P_{loc}(m|\rho_i)$, the probability that a site $i$ contains mass $m$ provided that the local density is $\rho_i = \langle m_i \rangle$. In the following calculations, the local mass distribution $P_{loc}(m|\rho_i)$ corresponding to a local density $\rho_i$ is replaced by the steady-state single-site mass distribution $P(m|\rho)$ calculated at density $\rho=\rho_i$. That is, we use an equality $P_{loc}(m|\rho) = P(m|\rho)$, which is expected to hold on the large spatio-temporal scales. Consequently, the steady-state mass distribution can be used to calculate also the other local observables, such as, the local occupation probability, which can be written as $\left\langle \hat{a}_i(t) \right\rangle_{\rho_i(t)=\rho} = a(\rho) = 1-P(m=0|\rho)$. For the notational simplicity, in the rest of the paper we denote the steady-state single-site mass distribution as $P(m) \equiv P(m|\rho)$.

In the presence of the small biasing force of magnitude $F \rightarrow 0$, the particle hopping rates change according to Eq.~(\ref{bias4}) and we obtain an exact expression of the density evolution equation,  
\begin{eqnarray}
\label{hydroKchip}
\frac{\partial \rho_i(t)}{\partial t} = \frac{1}{2} (g_{i-1} + g_{i+1} - 2 g_i) - \frac{F}{4} \delta x ( u_{i+1} - u_{i-1} ).~~~
\end{eqnarray}
Here the two local observables $g_i$ and $u_i$ are given by
\begin{eqnarray}
\label{g_i_general2}
g_i &=& p \left\langle \hat{a}_i \right\rangle + q \rho_i + q \sum_{v=0}^\infty \phi(v)   \left[ v \left\langle \hat{a}_i^v \right\rangle  - \left\langle m_i \hat{a}_i \hat{a}_i^v  \right\rangle \right], ~~~~
\\
\label{u_i_general2}
u_i &=& p \left\langle \hat{a}_i \right\rangle + q \left\langle m_i^2 \right\rangle + q \sum_{v=0}^\infty \phi(v) \left[ v^2 \left\langle \hat{a}_i^v \right\rangle -  \left\langle m_i^2 \hat{a}_i \hat{a}_i^v \right\rangle \right].~~~~ 
\end{eqnarray}
The calculation details for the derivation of Eq.~(\ref{hydroKchip}) are presented in Appendix \ref{app_1}. 
Note that the system satisfies ``gradient condition'' in the sense that the local diffusive current $J_D$ in Eq.~(\ref{hydroKchip}) can be written as a {\it gradient} (discrete) of a local observable $g_i$. The gradient property of these models is useful as it helps one to immediately identify the bulk-diffusion coefficient in the systems.
Now, by taking the diffusive scaling limit $i \rightarrow x = i/L$ and $t \rightarrow \tau = t/L^2$, and lattice constant $\delta x \rightarrow 1/L$,  Eq.~(\ref{hydroKchip}) leads to the time-evolution equation for the scaled coarse-grained density field $\rho(x, \tau) \equiv \rho_i(t)$,

\begin{eqnarray} 
\nonumber
\frac{\partial \rho(x,\tau)}{L^2 \partial \tau} &=& \frac{1}{2} \left[ g \left( x-\frac{1}{L}, \tau \right) + g \left( x + \frac{1}{L}, \tau \right) - 2g(x, \tau) \right] \\  \nonumber
&&- \frac{1}{4} \left[ u\left( x+\frac{1}{L}, \tau \right) - u\left( x - \frac{1}{L}, \tau \right) \right] \frac{F}{L}, \\
&=& \frac{1}{2} \left[ \frac{1}{L^2} \frac{\partial ^2 g(\rho)}{\partial x^2} \right] - \frac{1}{4} \left[ \frac{2}{L} \frac{\partial u(\rho)}{\partial x} \right] \frac{F}{L} +{\cal O} \left( \frac{1}{L^3} \right).~~~~~~
\label{grad-exp1}
\end{eqnarray}
Here we assume the density gradients being small ${\cal O}(1/L)$ and use the following small-gradient expansions,
\bea
 g \left( x \pm \frac{1}{L}, \tau \right) = g(\rho(x, \tau)) \pm \frac{1}{L} \frac{\partial g(\rho(x,\tau))}{\partial x}~~~~~~~~~~~~~~~~~~~~~~~~~\nonumber
\\ + \frac{1}{2 L^2} \frac{\partial^2 g(\rho(x, \tau))}{\partial x^2}
+ {\cal O} \left( \frac{1}{L^3} \right), ~~~~~
\label{grad-exp2}
\\
 u \left( x \pm \frac{1}{L}, \tau \right) = u(\rho(x, \tau)) \pm \frac{1}{L} \frac{\partial u(\rho(x, \tau))}{\partial x} + {\cal O} \left( \frac{1}{L^2} \right). ~~~~~~~
\label{grad-exp3}
\eea
Moreover, in the above expansions, we assume the existence of a local steady state, implying that, on the macroscopic spatio-temporal scales, the local quantities $g_i(t) \equiv g(\rho(x,\tau))$ and $u_i(t) \equiv u(\rho(x,\tau))$ depend on the coarse-grained (macroscopic) space and time variables $x$ and $\tau$, respectively, only through the coarse-grained density field $\rho(x,\tau)$. 
In the limit of $L \rightarrow \infty$,  Eq.~(\ref{grad-exp1}) immediately leads to the desired  hydrodynamic time-evolution equation of the density field,
\begin{eqnarray}
\label{hydroKchip2}
\frac{\partial \rho(x, \tau)}{\partial \tau} &=& \frac{1}{2} \frac{\partial ^2 g(\rho(x, \tau))}{\partial x^2} - \frac{F}{2}  \frac{\partial u(\rho(x, \tau))}{\partial x},
\end{eqnarray}
\noindent
where the quantities $g(\rho)$ and $u(\rho)$ can be calculated as a function of density $\rho$ from the single-site mass distribution as discussed below [see Eqs.~\eqref{g_i_general} and \eqref{u_i_general}]. Note that the above equation can be recast in the form of a continuity equation, 
\begin{equation}
\label{MFT_hydro}
\frac{\partial \rho(x,\tau)}{\partial \tau}=-\frac{\partial}{\partial x} \left[-D(\rho)\frac{\partial \rho}{\partial x} + \chi(\rho)F \right]  \equiv - \frac{\partial J(\rho)}{\partial x},
\end{equation}
through a constitutive relation between hydrodynamic current $J(\rho)$ and local density $\rho$,
 \be
J(\rho) = - D(\rho)\partial_x\rho + \chi(\rho)F,
\ee 
where  $D(\rho)$ and $\chi(\rho)$ are the bulk diffusion coefficient and the conductivity, respectively. The first term in the current arises according to Fick’s law where a nonuniform density profile contributes to a diffusive current $J_D (\rho) = - D(\rho)\partial_x \rho$ and the second term in the current provides a drift current $J_d (\rho) = \chi(\rho)F$, which is essentially the (linear) response to the small biasing force of magnitude $F$.
Comparing Eq.~(\ref{hydroKchip2}) with Eq.~(\ref{MFT_hydro}), one can identify the bulk-diffusion coefficient and conductivity, respectively as,
\begin{equation}
\label{transportKchip}
D(\rho)= \frac{1}{2} \frac{\partial g(\rho)}{\partial \rho}, ~~~~~ \chi (\rho) = \frac{u(\rho)}{2}.
\end{equation}
To explicitly calculate the transport coefficients $D(\rho)$ and $\chi(\rho)$ as a function of density $\rho$, we now use the identities $\left\langle \hat{a}_i \right\rangle = 1-P(0)$ and $\left\langle \hat{a}_i^{v} \right\rangle = \sum_{m=v}^{\infty}P(m)$ in Eq.~(\ref{g_i_general2}) and (\ref{u_i_general2}) and express $g(\rho)$ and $u(\rho)$ in terms of $P(m)$ as
\begin{eqnarray}
\label{g_i_general}
g(\rho_i) = \nonumber q \sum_{v=1}^\infty \phi(v) v \sum_{m=v}^\infty P(m) - q \sum_{v=2}^\infty \phi(v) \sum_{m=v}^\infty mP(m) \\  + p [1-P(0)] + q \rho_i \left[ 1-\phi(0) - \phi(1) \right], ~~~~~~
 \\
\label{u_i_general}
u(\rho_i) =  \nonumber q \sum_{v=1}^\infty \phi(v) v^2 \sum_{m=v}^\infty P(m) - q \sum_{v=2}^\infty \phi(v) \sum_{m=v}^\infty m^2P(m) \\ + p [1-P(0)] + q \left[ 1-\phi(0) - \phi(1) \right] \sum_{m=1}^\infty m^2 P(m). ~~~~~~
\end{eqnarray}
Note that the right hand sides of the above equations depend on the local density $\rho_i$ through the dependence of mass distribution $P(m)$ on the local density. So the task of calculating the transport coefficients essentially boils down to calculating the single-site mass distribution $P(m)$. Moreover, due to the gradient property, one would expect, through the macroscopic fluctuation theory \cite{Bertini_RMP2015}, the existence of an Einstein relation between the ratio of the two transport coefficients and the mass, or the particle-number, fluctuation in the systems,
\be
\frac{\chi(\rho)}{D(\rho)} = \sigma^2(\rho).
\label{ER}
\ee
Here the scaled subsystem mass fluctuation $\sigma^2(\rho)$ is defined as
\begin{equation}\label{sigma_sq}
\sigma^2(\rho) = \lim_{l_{sub} \rightarrow \infty}  \frac{\langle M_{sub}^2 \rangle - \langle M_{sub} \rangle^2}{l_{sub}},
\end{equation}
where $M_{sub} = \sum_{i=1}^{l_{sub}} m_i$ is the mass in a subsystem of size $l_{sub}$.

In the following sections, we explicitly calculate the local observables $g(\rho)$ and $u(\rho)$ as a function of density and demonstrate the above results obtained in Eqs.~(\ref{transportKchip}) and (\ref{ER}) for mass aggregation models with various choices of the probability distribution $\phi(v)$.
Note that, unless mentioned otherwise, we take $p=q=1/2$ throughout; extension of the results to generic values of $p$ and $q$ is straightforward.

\section{Variant I: Fragmentation of a fixed amount of mass}
\label{Sec_VarI}

Depending on the choice of the probability distribution function $\phi(v)$, we consider several special cases of the generalized model described in Sec.~\ref{Model}, some of which can be solved analytically. The special cases we discuss in this section have a sharply localized distribution $\phi(v)=\delta_{v,v_0}$. 
In order to calculate the transport coefficients, we need to determine two local observables $g(\rho)$ and $u(\rho)$ as a function of density $\rho$. So by putting $\phi(v)=\delta_{v,v_0}$ and setting $p=q=1/2$ in Eqs.~(\ref{g_i_general2}) and (\ref{u_i_general2}), the local observables are written in a simplified form,
\begin{eqnarray}
\label{local_g}
g(\rho_i) = \frac{1}{2} \left[ \rho_i+\left\langle \hat{a}_i \right\rangle + v_0\left\langle \hat{a}_i^{v_0} \right\rangle - \left\langle m_i \hat{a}_i \hat{a}_i^{v_0} \right\rangle \right], ~~~~~\\
\label{local_u}
u(\rho_i) = \frac{1}{2} \left[ \left\langle \hat{a}_i \right\rangle - \left\langle m_i^2 \hat{a}_i \hat{a}_i^{v_0} \right\rangle + \left\langle m_i^2 \right\rangle + {v_0}^2 \left\langle \hat{a}_i^{v_0} \right\rangle \right].~~~~
\end{eqnarray}
In the following sections, we explore a few explicitly solvable cases having three different values of $v_0=1$, $2$ and $\infty$; as demonstrated later, the model with $v_0=\infty$, which was previously studied in Ref. \cite{Barma_PRL1998, Barma_JSP2000}, undergoes a condensation transition upon tuning the global density of the system. However, for generic values of $v_0$, the transport coefficients are calculated numerically using the above equations.

\subsection{Case I: $v_0=1$, zero range process }
\label{Sec_ZRP}

In this section, we  illustrate the general hydrodynamic formalism developed in the previous section by starting with the simplest case: variant I with $\phi(v)=\delta_{v,1}$. Note that this particular case is equivalent to that with pure single-particle chipping, or with pure single-particle aggregation, occurring with rate $(p+q)$. As we see later, the steady-state probability weights of microscopic configurations and the hydrodynamic time-evolution equations, up to a trivial rescaling of time,  are independent of the rates $p$ and $q$. The mass $m_i(t)$ at site $i$ and at time $t$ for the biased system is updated in an infinitesimal time interval  $dt$ as given below,
\begin{eqnarray} \nonumber
m_i(t+dt) =~~~~~~~~~~~~~~~~~~~~~~~~~~~~~~~~~~~~~~~~~~\\
\left\{
\begin{array}{ll}
m_i(t) - 1            & {\rm prob.}~ p \hat{a}_i dt, \\
m_i(t) + 1            & {\rm prob.}~ p\hat{a}_{i-1} \left(1+\frac{F\delta x}{2}\right) \frac{dt}{2}, \\
m_i(t) + 1            & {\rm prob.}~ p\hat{a}_{i+1} \left(1-\frac{F\delta x}{2}\right) \frac{dt}{2}, \\
m_i(t) - 1            & {\rm prob.}~ q \hat{a}_i dt, \\
m_i(t) + 1            & {\rm prob.}~ q\hat{a}_{i-1} \left(1+\frac{F\delta x}{2}\right) \frac{dt}{2}, \\
m_i(t) + 1            & {\rm prob.}~ q\hat{a}_{i+1} \left(1-\frac{F\delta x}{2}\right) \frac{dt}{2}, \\
m_i(t)                & {\rm prob.}~ 1-\Sigma dt,
\end{array}
\right.
\label{EqZRP}
\end{eqnarray}
with,
\begin{eqnarray}
\Sigma = (p+q)\left[\hat{a}_i + \frac{1}{2}\left\{\hat{a}_{i-1}\left(1+\frac{F\delta x}{2}\right)+ \hat{a}_{i+1}\left(1-\frac{F\delta x}{2}\right)\right\}\right]. \nonumber \\
\end{eqnarray}
 The local density $\rho_i(t) = \left\langle m_i(t) \right\rangle$ at site $i$ and time $t$ evolves as
\begin{eqnarray} 
\frac{\partial \rho_i}{\partial t} = \frac{(p+q)}{2}[\left(\left\langle \hat{a}_{i+1}\right\rangle + \left\langle \hat{a}_{i-1}\right\rangle - 2 \left\langle \hat{a}_{i}\right\rangle\right) \nonumber \\ +\frac{F\delta x}{2}\left( \left\langle\hat{a}_{i-1}\right\rangle-\left\langle\hat{a}_{i+1}\right\rangle \right)].
\end{eqnarray}
Now in the diffusive scaling limit, $i \rightarrow x = i/L$ and $t \rightarrow \tau = t/L^2$, and lattice constant $\delta x \rightarrow 1/L$, the above equation leads to the hydrodynamic evolution of density field $\rho(x,\tau)$, 
\begin{eqnarray}
\label{hydroZRP}
\frac{\partial \rho(x, \tau)}{\partial \tau} &=& \frac{(p+q)}{2} \frac{\partial ^2 a(\rho)}{\partial x^2} - \frac{(p+q)}{2} F \frac{\partial a(\rho)}{\partial x},
\end{eqnarray}
where,
 \begin{equation} \label{occupation_prob}
 a(\rho) = 1-P(m=0),
\end{equation}  
is the probability that a site is occupied.
By comparing Eq.~(\ref{hydroZRP}) with Eq.~(\ref{MFT_hydro}), the bulk diffusion coefficient and the conductivity can be readily identified as
\begin{eqnarray}
\label{ZRP_diff1}
D(\rho)&=&\frac{(p+q)}{2}\frac{\partial a(\rho)}{\partial \rho}, \\
\label{ZRP_cond1}
 \chi(\rho)&=&\frac{(p+q)}{2}a(\rho).
\end{eqnarray}
To explicitly calculate the transport coefficients as a function of density, we first try to obtain the single-site mass distribution $P(m)$, which can be calculated using the steady-state joint mass-distribution 
\be
{\cal P}(m_1, m_2, \dots, m_L)= \prod_{k=1}^{L} P(m_k),
\label{joint-dis}
\ee 
which, in this case, has a product form. The above product form can be easily understood from the fact that the unbiased process, i.e., Eq.~(\ref{EqZRP}) with $F=0$, is a zero range process (ZRP) with particle-hopping rates being constant \cite{Hanney_Evans_JPhysA}. Indeed, as we demonstrate in Appendix \ref{app_3},  the neighboring spatial correlations vanish in the mass aggregation processes for generic parameter values so that one can in principle resort to a mean-field analysis, similar to the one performed below.

For completeness, we now present a derivation of the single-site mass distribution $P(m)$ for $v_0=1$ using a master equation method along the lines of Ref. \cite{Puri}. Indeed the analysis provided below illustrates our overall strategy in calculating the transport coefficients in various other cases discussed later. The time evolution of the single-site mass distribution can be written as
\begin{eqnarray}
 \nonumber
\frac{dP(m,t)}{dt} &=& (p+q) [-P(m, t) -a(\rho)P(m, t) +P(m+1, t) \\
\label{ZRP_Meq1}
&&+a(\rho)P(m-1, t)],~~~\text{for}~m>0, \\
\label{ZRP_Meq2}
\frac{dP(0,t)}{dt} &=& (p+q)\left[-a(\rho)P(0, t) +P(1, t)\right].
\end{eqnarray}
In the steady state, Eq.~(\ref{ZRP_Meq2}) provides a condition
\be
\label{ZRP_Meq3}
P(1)=aP(0)=P(0)[1-P(0)].
\ee
Now by defining the steady-state generating function  $Q(z)=\sum_{m=1}^{\infty} P(m)z^m$, multiplying Eq.~(\ref{ZRP_Meq1}) by $z^m$ and then summing over $m$ from $1$ to $\infty$, we obtain
\bea
\label{ZRP_Meq4}
-(1+a)Q(z) +\frac{1}{z}[Q(z)-zP(1)]+az[Q(z)+P(0)]=0,~~~~~~
\eea
which, after substituting Eq.~(\ref{ZRP_Meq3}) into Eq.~(\ref{ZRP_Meq4}), leads to
\bea
\label{ZRP_Meq5}
Q(z)=\frac{zP(0)[1-P(0)]}{1-z[1-P(0)]}.
\eea
To determine $P(0)$, we use the condition $\frac{dQ(z)}{dz}\big\vert_{z=1} = \left\langle m \right\rangle= \rho$ to obtain
\be
\label{ZRP_P0}
P(0)=\frac{1}{1+\rho}.
\ee
After substituting Eq.~(\ref{ZRP_P0}) into Eq.~(\ref{ZRP_Meq5}) and expanding $Q(z)$ in powers of $z$,
\bea
Q(z) = \frac{1}{1+\rho}\sum_{m=1}^{\infty}\left( \frac{z\rho}{1+\rho} \right)^m,
\eea
we obtain exactly the steady-state mass distribution,
\bea
\label{ZRP_dist}
P(m)=\frac{1}{1+\rho} \left( \frac{\rho}{1+\rho} \right)^m.
\eea
Now the analytic expression of occupancy,
\be
a(\rho)=\frac{\rho}{1+\rho},
\ee
is used in Eqs.~(\ref{ZRP_diff1}) and (\ref{ZRP_cond1}) to finally obtain the bulk-diffusion coefficient $D(\rho)$ and the conductivity $\chi(\rho)$ as a function of density,
\bea
\label{ZRP_diff2}
D(\rho)&=&\frac{(p+q)}{2(1+\rho)^2}, \\
\label{ZRP_cond2}
\chi(\rho)&=&\frac{(p+q)\rho}{2(1+\rho)}.
\eea
One can immediately check the Einstein relation Eq.~(\ref{ER}) by directly calculating the scaled subsystem mass fluctuation as
\be
\label{ZRP_fluctuation}
\sigma^2(\rho) \equiv \lim_{l_{sub} \rightarrow \infty} \frac{\langle M_{sub}^2 \rangle - \langle M_{sub} \rangle^2}{l_{sub}}  = \rho(1+\rho),
\ee
using the fact that $\langle M_{sub}^2 \rangle - \langle M_{sub} \rangle^2 = l_{sub} (\langle m^2 \rangle - \langle m \rangle^2)$  as the neighboring correlations, in the limit of large system size, identically vanish, i.e., $c(r) = \langle m_i m_{i+r} \rangle - \rho^2 = 0$ for $r \ne 0$, due to the steady-state product measure in Eq.~(\ref{joint-dis}).

\subsection{Case II: $v_0=2$} 
\label{Sec_v2}

We now consider the first non-trivial case, that is variant I with $v_0=2$; this model can be mapped to an exclusion process with nearest and next-nearest-neighbor particle hopping \cite{Hanney_Evans_JPhysA}. As the neighboring correlations are shown to vanish in the limit of large system sizes (see Appendix \ref{app_3}), we can calculate the steady-state single-site mass distribution $P(m)$ by employing a mean-field theory, where the joint mass distribution is assumed to have a product form. Now taking into account all possible ways of mass transfer, we can write the time evolution equations of $P(m,t)$ for an arbitrary $v_0$ as given below,
\begin{widetext}
\begin{eqnarray} \nonumber
\label{m1}
&&\frac{d P(m,t)}{dt} \bigg\lvert_{m>0} = - (p+q) \left( 1 + \sum_{m'=1}^{\infty} P(m',t) \right) P(m,t) + q P(m+v_0,t) + p P(m+1,t) + p P(m -1,t) \sum_{m'=1}^{\infty} P(m',t)  \\ 
&& + q P(m-v_0,t) \Theta(m-v_0) \sum_{m'=v_0}^{\infty}P(m',t) + q \sum_{m'=1}^m P(m-m',t) P(m',t) - q \sum_{m'=v_0}^m P(m-m',t) P(m',t) \Theta(m-v_0),~~ \\
\label{m0}
&&\frac{d P(0,t)}{dt} = -(p+q) \sum_{m'=1}^{\infty} P(m',t) P(0,t)+ p P(1,t) + q \sum_{m'=1}^{\infty} P(m',t) + q P(v_0,t) - q \sum_{m'=v_0}^{\infty} P(m',t),
\end{eqnarray} 
\end{widetext}
where the Heaviside step function $\Theta(m-v_0) = 0$ if $m < v_0$ and $\Theta(m-v_0)= 1$ otherwise. We now solve the master equations (\ref{m1}) and (\ref{m0}) for a particular value of $v_0=2$ in the steady state by setting the left hand sides of Eqs.~(\ref{m1}) and (\ref{m0}) to zero. Now multiplying the right hand side of Eq.~(\ref{m1}) by $z^m$, summing $m$ from $1$ to $\infty$, and by combining Eq.~(\ref{m0}) in the steady state, we solve for the generating function $Q(z)= \sum_{m=1}^{\infty} P(m) z^m$ as given below,
\begin{eqnarray}
\label{Gen_fn_p_q}
Q(z)= \frac{z\left[\tilde{q}P_1+P_0(1-P_0)(1+\tilde{q}z)z-\tilde{q}P_1P_0 z^2\right]}{\tilde{q}+\{z-(1-P_0)z^{2}\}-\tilde{q}(1-P_1-P_0) z^3 },~~~~
\end{eqnarray} 
where we write $\tilde{q}=q/(p+q)$. We further simplify the problem by choosing $p=q=1/2$ and in this case we obtain,
\begin{eqnarray}
\label{Gen_fn}
Q(z)= z\frac{P_1+P_0(1-P_0)(2+z)z-P_1P_0 z^2}{1+2z-2(1-P_0)z^{2}-(1-P_1-P_0) z^3 },~~~
\end{eqnarray} 
where we denote the undetermined parameters $P_0=P(m=0)$ and $P_1=P(m=1)$. By definition, we have $Q(0)=0$ and $Q(1)=1-P_0$, both of which are satisfied by Eq.~(\ref{Gen_fn}), implying that the above expression for $Q(z)$ is indeed consistent. To determine the two unknown parameters $P_0$ and $P_1$ in the generating function $Q(z)$, we need to put two conditions on $Q(z)$. One condition can be found from the identity $\frac{dQ}{dz}\big\vert_{z=1} = \rho$, which leads to
\be
\label{Con1}
P_1=\frac{5-P_0(5+3\rho)}{\rho+2}.
\ee
The second condition is obtained as follows. From the definition, $Q(z)$ converges only if $|z| \leq 1$ since $0\le P(m)\le 1$. However, if the denominator of $Q(z)$ has a root at $z=z^*$ with $|z^*| \leq 1$, $Q(z)$ will diverge at that root $z^*$ which is not allowed. So to avoid a diverging $Q(z)$, both the denominator and the numerator of $Q(z)$ in Eq.~(\ref{Gen_fn}) should share a common root at $z=z^*$ so that $Q(z)$ remains finite. This condition helps us to determine the probability $P_1$ in terms of probability $P_0$ and density $\rho$. As the numerator of $Q(z)$ is a quadratic function of $z$, we explicitly find the two roots,
\begin{equation}
\label{roots}
z_{\pm}=\frac{1-P_0}{1-P_1-P_0}\left[-1\pm\sqrt{1-\frac{P_1(1-P_1-P_0)}{P_0(1-P_0)^{2}}}\right].
\end{equation}

\begin{figure}[h]
\begin{center}
\includegraphics[width=9cm,angle=0]{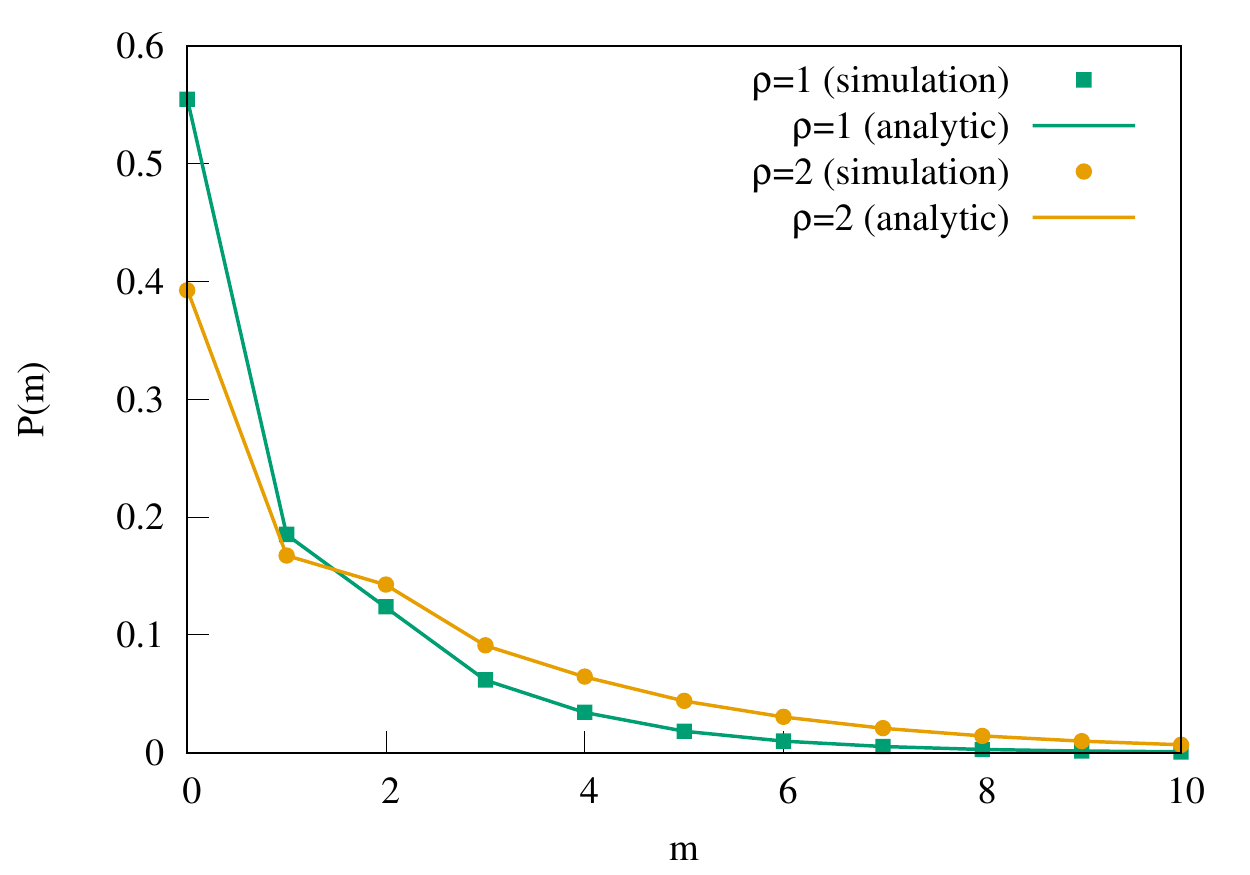}
\caption{ {\it Variant I, $v_0=2$.} Single-site mass distributions $P(m)$ are plotted as a function of mass $m$ for densities $\rho=1$ (green squares) and $2$ (yellow circles). Simulations (points) and the exact mean-field theoretical results (lines) as in Eq.~(\ref{distn_v2}) are in excellent agreement with each other.}
\label{m_dist_v2}
\end{center}
\end{figure}

Since $0 < P_1,~P_0 < 1$, the pre-factor ${(1-P_0)}/{(1-P_1-P_0)}$ in the above equation is always greater than $1$. Moreover, one can check that the term inside the square root is always positive, implying that both the roots are real and $z_{-} \leq -1$. Therefore the root of physical interest is $z=z_{+}$. Furthermore, the denominator of $Q(z)$ in Eq.~(\ref{Gen_fn}) should vanish at $z=z^*=z_{+}$. Using this condition and the relation in Eq.~(\ref{Con1}) together, we express the probabilities $P_0$ and $P_1$ as a function of density $\rho$,
\bea
\label{P0}
P_0 = \frac{9+5\rho-\sqrt{1+10\rho+5\rho^{2}}}{2(2+\rho)^{2}}, ~~~~~~~~~~~~~~~~~~~~~~ \\
\label{P1}
P_1 = \frac{(3\rho+5)\sqrt{1+10\rho+5\rho^{2}}-(5\rho^{2}+12\rho+5)}{2(2+\rho)^{3}}.~~
\eea
Next we expand the generating function $Q(z)$ as Eq.~(\ref{Gen_fn}) in power series of $z$,
\be
\label{Gen_fn_v2}
Q(z)=\sum_{m=1}^{\infty} \left( \frac{P_1}{P_0} \right)^m P_0 F_{m+1} z^m,
\ee
where $F_{m+1}$ is the $(m+1)$th element of the Fibonacci sequence, where $m$th element is defined as the sum of the two preceding ones,
\begin{equation}\label{Fibonacci}
F_m= F_{m-1} + F_{m-2},
\end{equation}
 for $m \geq 2$ and the first two terms are given by  $F_0=0$ and $F_1=1$ \cite{beck2010art}.
 Comparing the above power series expansion and the definition of the generating function $Q(z)$, we immediately find the singe-site mass distribution $P(m)$ as a function of $m$ for any density $\rho$,
\be
\label{distn_v2}
P(m)=\left( \frac{P_1}{P_0} \right)^m P_0 F_{m+1},
\ee
where $P_0$ and $P_1$ both depend on density and are provided by Eqs.~(\ref{P0}) and (\ref{P1}), respectively. One can  show that variant I with $v_0=2$ (any other $v_0$, except $v_0=1$) violates detailed balance  and the joint-mass distributions cannot be written in terms of the equilibrium Boltzmann  distribution (See Appendix \ref{app_2}). In Fig. \ref{m_dist_v2}, we plot, for two different densities $\rho=1$ (green squares) and $2$ (yellow circles), the single-site mass distributions $P(m)$ as a function of mass $m$, obtained from simulations, which are in excellent agreement with the analytic expression as in Eq.~(\ref{distn_v2}).

Using the mean-field analysis similar to that performed above, it is in principle possible to find the steady-state single-site mass distributions, and therefore to obtain the transport coefficients, also for other values of $v_0$. However, the calculations are outside the scope of the present work.

\subsubsection{Transport coefficients and density relaxation}
\label{Sec_Spreadingv2}

To explicitly calculate the transport coefficients as a function of density, one needs to first evaluate the two density-dependent local observables $g(\rho)$ and $u(\rho)$. This  can be done by directly calculating the steady-state averages in Eqs.~\eqref{g_i_general} and \eqref{u_i_general} with the help of the mass distribution $P(m)$ in Eq.~(\ref{distn_v2}). Now, by substituting $g(\rho)$ and $u(\rho)$ in Eq.~\eqref{transportKchip} and performing somewhat tedious but straightforward algebraic manipulations, we explicitly find the analytic expressions of the bulk-diffusion coefficient and the conductivity, 
\begin{widetext}
\begin{eqnarray}
\label{D_v2}
&& D(\rho)=\frac{1}{4} \frac{5\rho \left(2\rho \left(\sqrt{5\rho (2+\rho)+1}-1\right)+10 \sqrt{5 \rho (2+\rho)+1}-3\right)+57 \sqrt{5 \rho (2+\rho)+1}+7}{2 (2+\rho)^4 \sqrt{5\rho (2+\rho)+1}}, \label{D_v2}
\\
\label{Chi_v2}
&& \chi(\rho) = \frac{1}{4} \frac{-5 \sqrt{5 \rho (2+\rho)+1} + \rho \left(10 \rho (5+\rho)-4 \sqrt{5 \rho (2+\rho)+1}+61\right)+5}{2 (2+\rho)^3}.
\end{eqnarray}
\end{widetext}

\begin{figure}[h]
\begin{center}
\leavevmode
\includegraphics[width=9cm,angle=0]{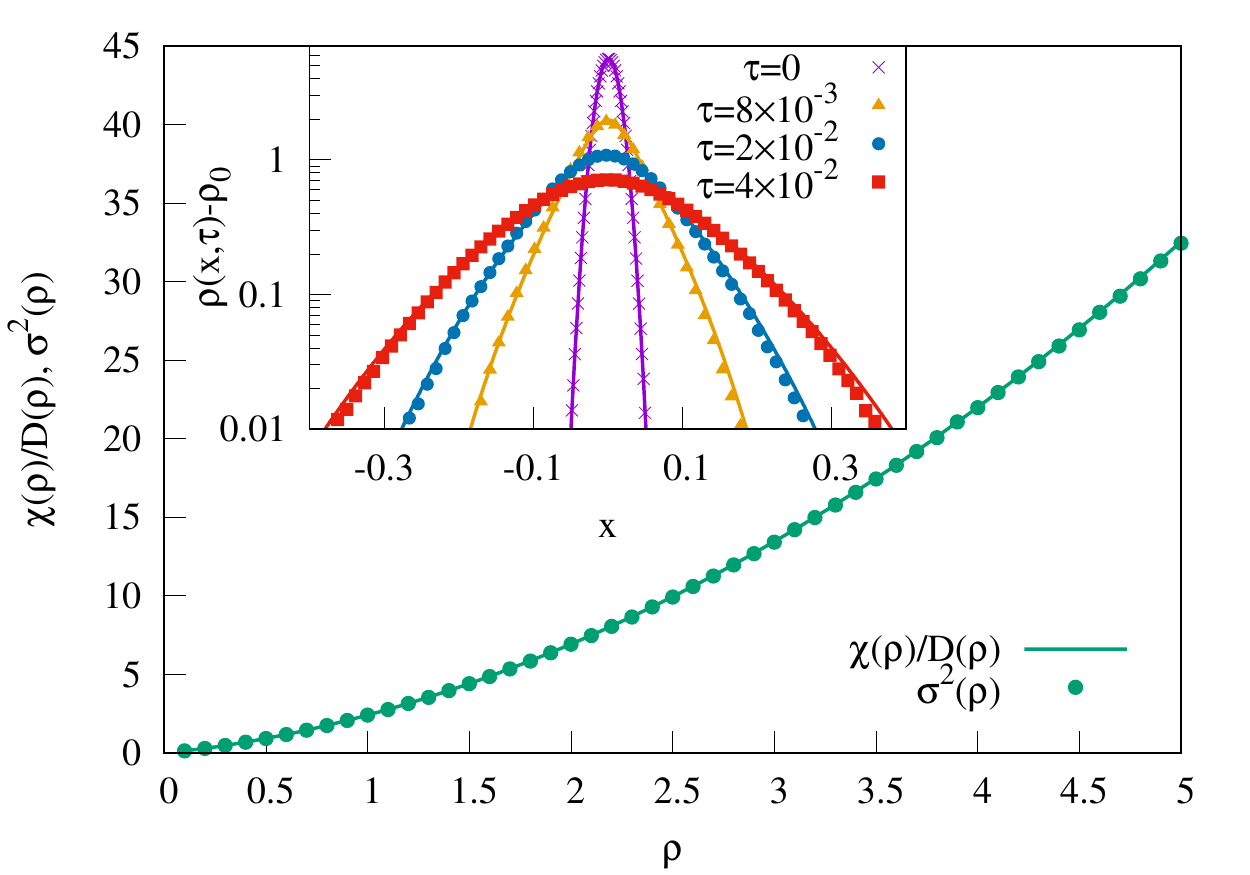}
\caption{ {\it Verification of Einstein relation and density relaxation in Variant I, $v_0=2$}: We verify the Einstein relation by plotting scaled mass fluctuation $\sigma^2(\rho)$, calculated from simulations (green circle), and compare that with the analytical expression (green line) of  the ratio of the transport coefficients ${\chi(\rho)}/{D(\rho)}$ obtained from Eqs.~(\ref{D_v2}) and (\ref{Chi_v2}). In inset, we verify density relaxation of an initially localized density perturbation [Eq.~(\ref{initialKchip})] at different hydrodynamic times $\tau=0$ (magenta cross), $8\times 10^{-3}$ (yellow triangle),  $2\times 10^{-2}$ (blue circle), and $4\times 10^{-2}$ (red square); we take $\rho_0=0.5$, $n_1=0.2$ and $\Delta^2 = 2\times10^{-4}$ in the initial density profile as in Eq.~(\ref{initialKchip}). The lines denote numerically integrated hydrodynamic time-evolution obtained by using the functional form of $D(\rho)$ as in Eq.~\eqref{D_v2} and points denote simulation results.}
\label{ERv2}
\end{center}
\end{figure}

First we verify the above expression of the bulk-diffusion coefficient $D(\rho)$ by studying the relaxation of an initial density perturbation. To this end, we numerically integrate the nonlinear diffusion equation (\ref{MFT_hydro}) with external biasing force $F=0$, 
\begin{equation}
\label{hydroKchip5}
\frac{\partial \rho(x, \tau)}{\partial \tau}=\frac{\partial}{\partial x} \left[D(\rho)\frac{\partial \rho}{\partial x} \right].
\end{equation}
We start by considering an initial density profile,

\begin{equation}
\label{initialKchip}
\rho(x, \tau =0) = \rho_0 + n_1 \frac{\exp(-x^2/2 \Delta^2)}{\sqrt{2 \pi \Delta^2}},
\end{equation}

where the background density $\rho_0$ is uniform, $n_1$ is the strength of the density perturbation due to the addition of excess particles over the uniform background, and $\Delta$ is the width (or the standard deviation) of the initial density perturbation. We choose  $\Delta^2 = 2\times10^{-4}$ and $n_1=0.2$. For the numerical integration, we employ the  Euler integration scheme, where we discretize the variables $x$ and $\tau$ in Eq.~(\ref{hydroKchip5}).
We also perform Monte Carlo simulations of the mass aggregation processes by employing random sequential updates (which corresponds to the continuous-time dynamics) and the same initial condition as in Eq.~(\ref{initialKchip}). 
In the simulations, the averaging was done over various initial configurations as well as the trajectories.
In inset of Fig. \ref{ERv2}, we compare the density profiles obtained by numerically integrating the hydrodynamic evolution Eq.~(\ref{hydroKchip5}) and that obtained from simulations, at various hydrodynamic times $\tau=0$ (magenta cross), $8\times 10^{-3}$ (yellow triangle),  $2\times 10^{-2}$ (blue circle), and $4\times 10^{-2}$ (red square).  The hydrodynamic theory (lines) captures the simulation results (points)  reasonably well over several decades of density values.

We also verify the Einstein relation as given in Eq.~(\ref{ER}). For this purpose,  we directly compute, from the knowledge of the single-site mass distribution in Eq.~(\ref{distn_v2}), the scaled variance $\sigma^2(\rho) = \lim_{l_{sub} \rightarrow \infty} [\langle M_{sub}^{2} \rangle - \langle M_{sub} \rangle^2]/{l_{sub}}$ of subsystem mass $M_{sub}$ in a large subsystem of size $l_{sub}$. The scaled variance $\sigma^2(\rho)$, within the mean-field theory (as verified in Appendix \ref{app_2}), is exactly equal to the steady-state variance  of single-site mass $m$, and is given by
\bea
 \sigma^2(\rho) &=& \langle m^2 \rangle - \langle m \rangle^2, 
\nonumber \\
  &=& \rho^2+2 \rho -\frac{1}{5} \sqrt{5 \rho (\rho+2)+1}+\frac{1}{5}. 
\label{var_v02}
\eea
After some algebraic manipulations using Eqs.~(\ref{D_v2}) and (\ref{Chi_v2}), it can indeed be shown that the expression of the scaled mass fluctuation $\sigma^2(\rho)$ given in Eq.~(\ref{var_v02})
is exactly the same as the ratio of the two transport coefficients ${\chi(\rho)}/{D(\rho)}$, immediately implying the Einstein relation Eq.~(\ref{ER}). In Fig. \ref{ERv2}, we plot the scaled mass fluctuation $\sigma^2(\rho)$ obtained from simulations (circles) and compare the simulation results with the analytical expression of the ratio ${\chi(\rho)}/{D(\rho)}$ (lines), obtained from Eqs.~(\ref{D_v2}) and (\ref{Chi_v2}); the agreement between theory and simulations is excellent.

Now we discuss the behaviors of the transport coefficients in the two limiting cases of small and large densities. In the low density limit $\rho \rightarrow 0$, the probability of a site occupied by two or more masses is very small. In this limit, bulk-diffusion coefficient in Eq.~(\ref{D_v2}) becomes $D(\rho)\simeq 1/2 + {\cal O}(\rho)$ and the conductivity given by Eq.~(\ref{Chi_v2}) becomes $\chi(\rho)\simeq \rho/2 + {\cal O}(\rho^2)$, resulting in the mass fluctuation $\sigma^{2}(\rho)=\chi/D \simeq \rho$ in the leading order of density. This is expected as, in the low density limit, the mass distribution is given by the Poisson distribution for which the fluctuation is equal to the mean. In the other limit of high density $\rho \rightarrow \infty$, $D(\rho)\simeq 5/4 \rho^2$ and $\chi(\rho)\simeq 5/4$, and thus the fluctuation $\sigma^{2}(\rho)=\chi/D\simeq \rho^{2}$, which is the same as the large density behavior of the scaled subsystem mass fluctuation in the ZRP [see Eq.~(\ref{ZRP_fluctuation})].

\subsection{Case III: $v_0 \rightarrow \infty$}
\label{Sec_v_inf}

In this section, we consider the most interesting case of infinitely large $v_0 \rightarrow \infty$. This case was studied  in Refs. \cite{Barma_PRL1998, Barma_JSP2000} to understand the steady-state properties of clustering phenomena in mass aggregation processes. The model allows for single-particle chipping, diffusion and aggregation of masses.
Beyond a critical global density $\rho > \rho_c$, a condensation transition was observed with a macroscopic-size mass aggregate forming in the system, in addition to the power-law single-site mass distribution in the bulk. In this section, we calculate the bulk-diffusion coefficient and the conductivity and characterize the condensation transition in the light of an underlying instability, or a singularity, in the conductivity. To this end, we resort to a mean-field theory, which helps us to obtain the explicit expressions of the transport coefficients.

As $v_0 \rightarrow \infty$, mass $m_i$ at a site $i$ is always less than $v_0$, implying that the indicator function $a_i^v$ is zero. By putting $a_i^v=0$ in Eq.~(\ref{General_Update}), the continuous-time update for mass at site $i$ in infinitesimal time interval can be written as
\begin{eqnarray} \nonumber
m_i(t+dt) = ~~~~~~~~~~~~~~~~~~~~~~~~~~~~~~~~~~~~~~~~~~~~~~~~~~~~~\\
\left\{
\begin{array}{ll}
0                     & {\rm prob.}~ \hat{a}_i dt/2, \\
m_i(t) - 1            & {\rm prob.}~ \hat{a}_i dt/2, \\
m_i(t) + 1            & {\rm prob.}~ \hat{a}_{i-1} dt/4, \\
m_i(t) + 1            & {\rm prob.}~ \hat{a}_{i+1} dt/4, \\
m_i(t) + m_{i-1}(t)   & {\rm prob.}~ \hat{a}_{i-1} dt/4, \\
m_i(t) + m_{i+1}(t)   & {\rm prob.}~ \hat{a}_{i+1} dt/4, \\
m_i(t)                & {\rm prob.}~ 1-[\hat{a}_i + \frac{1}{2}(\hat{a}_{i-1}+ \hat{a}_{i+1})] dt .
\end{array}
\right.
\label{Update_unbiased_Barma}
\end{eqnarray}
The above update equation can be used to obtain the following equation for the second moment of mass $\left\langle m_i^2(t) \right\rangle$ in the steady state,
\begin{align}
\left\langle m_i^2 \right\rangle =& \frac{1}{4} \left[ \left\langle 2(m_i-1)^2\hat{a}_i \right\rangle + \left\langle (m_i+1)^2\left(\hat{a}_{i-1}+\hat{a}_{i+1}\right) \right\rangle \right. \nonumber \\ 
& + \left\langle (m_i + m_{i-1})^2 \hat{a}_{i-1} \right\rangle + \left\langle (m_i + m_{i+1})^2 \hat{a}_{i+1} \right\rangle \nonumber \\
& \left. + 4\left\langle m_i^2 \right\rangle - \left\langle m_i^2 \left\{4\hat{a}_i + 2(\hat{a}_{i-1}+ \hat{a}_{i+1})\right\} \right\rangle \right] , \label{second_moment_Barma}
\end{align}
where we have used the steady-state condition $\left\langle m_i^2(t+dt) \right\rangle = \left\langle m_i^2(t) \right\rangle$. 
As demonstrated in Appendix \ref{app_3}, the finite-size scaling analysis of the two-point spatial correlation functions implies that the neighboring correlations vanish as $L \rightarrow \infty$. Therefore the two-point correlation functions in Eq.~(\ref{second_moment_Barma}) can be written as a product of one-point functions, such as $\langle m_i^2 m_j \rangle = \langle m_i^2 \rangle \langle m_j \rangle$, $\langle m_i^2 \hat a_j \rangle = \langle m_i^2 \rangle \langle \hat a_j \rangle$, and $\langle m_i m_j^2 \hat a_j \rangle = \langle m_i \rangle \langle m_j^2 \hat a_j \rangle$, etc. Then using the identities, $\langle m_j^2 \hat a_j \rangle = \langle m_j^2 \rangle$ and $\langle m_j \hat a_j \rangle = \rho$, we straightforwardly obtain the occupation probability,
\be
a(\rho) = \frac{\rho (1-\rho)}{1+\rho},
\label{occupancy_Barma}
\ee
 as a function of density. Similarly, by using the steady state condition for the third moment of local mass $\left\langle m_i^3(t+dt) \right\rangle = \left\langle m_i^3(t) \right\rangle$, we calculate, exactly within the mean-field theory, the second moment $\left\langle m_i^2 \right\rangle \equiv \theta_2(\rho)$,
\be
\theta_2 (\rho) = \frac{\rho [1+a(\rho)]}{1-a(\rho) - 2 \rho}.
\label{third_moment_Barma}
\ee
After substituting Eq.~(\ref{occupancy_Barma}) into the above equation, we readily obtain the scaled subsystem mass fluctuation,
\begin{eqnarray} \nonumber
\lim_{l_{sub} \rightarrow \infty} \frac{\langle M_{sub}^{2} \rangle - \langle M_{sub} \rangle^2}{l_{sub}} \equiv \sigma^2(\rho)\\
 = \theta_2(\rho) - \rho^2 = \frac{\rho (1+ \rho) (1+\rho^2)}{1 - 2 \rho - \rho^2}.
\label{fluctuation_Barma}
\end{eqnarray}
Note that the mass fluctuation diverges beyond a critical density $\rho_c=\sqrt{2}-1$ and signals a condensation transition \cite{Das_PRE2015}, which was previously observed in Refs. \cite{Barma_PRL1998, Barma_JSP2000} in this particular variant of  the mass aggregation process.

\subsubsection{Transport coefficients and density relaxation}
\label{Sec_v_inf_ER}

\begin{figure}
\begin{center}
\includegraphics[width=9cm,angle=0]{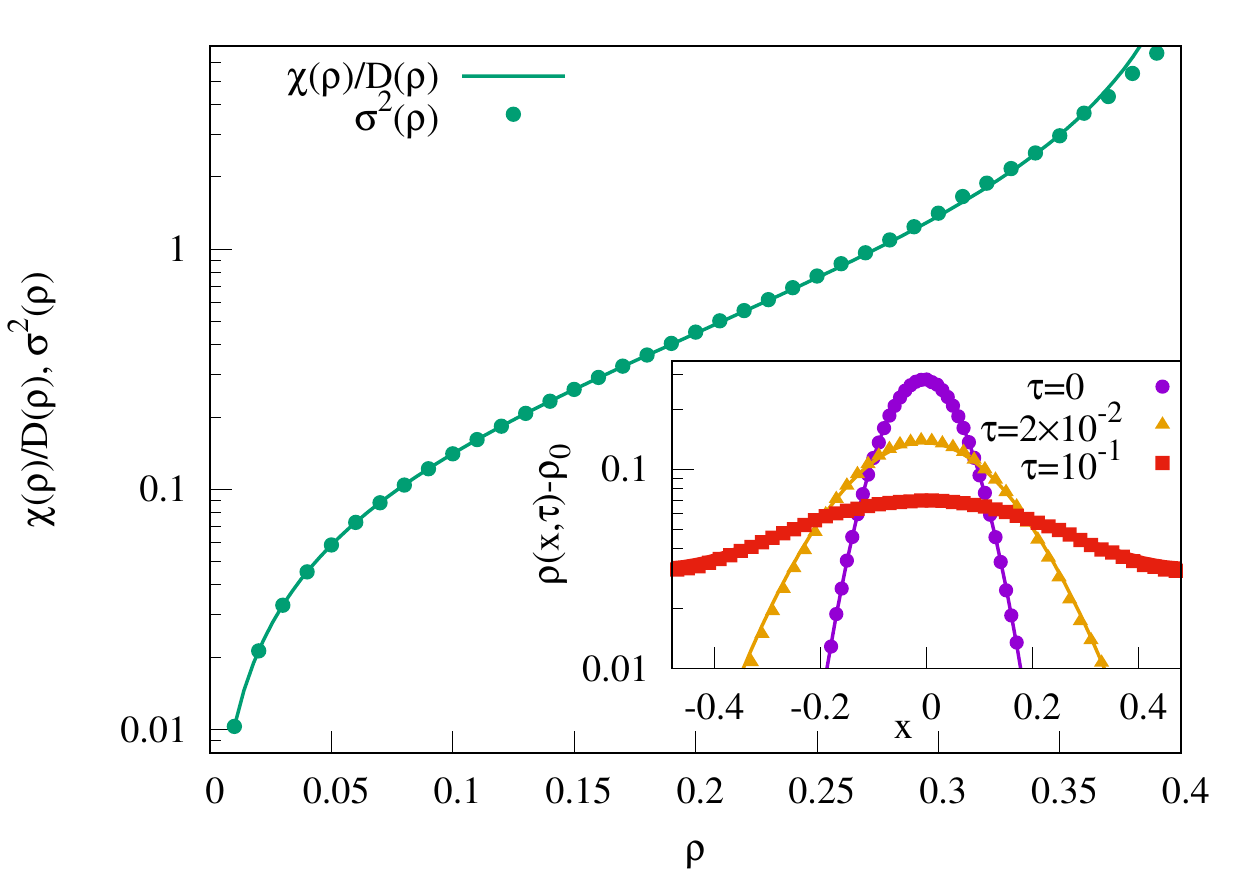}
\caption{ {\it Variant I, $v_0 \rightarrow \infty$}: We verify the Einstein relation by plotting scaled subsystem mass fluctuation $\sigma^2(\rho)$, calculated from simulations (green circle), and compare that with the analytical expression (green line) of the ratio of the transport coefficients ${\chi(\rho)}/{D(\rho)}$ obtained from Eqs.~(\ref{diffusivityBarma}) and (\ref{conductivityBarma}). In inset, we verify density relaxation at various final hydrodynamic times $\tau= 0$ (magenta circle), $2\times 10^{-2}$ (yellow triangle), and $10^{-1}$ (red square). We take the initial density perturbation as in Eq.~(\ref{initialKchip}) with $\rho_0=0.1$, $n_1=0.05$ and $\Delta^2=5\times 10^{-3}$. The lines denote numerically integrated hydrodynamic time-evolution obtained using the functional form of $D(\rho)$ as in Eq. \eqref{diffusivityBarma} and points denote simulation results.}
\label{FigHydroBM}
\end{center}
\end{figure}

\begin{figure}[h]
\begin{center}
\leavevmode
\includegraphics[width=9cm,angle=0]{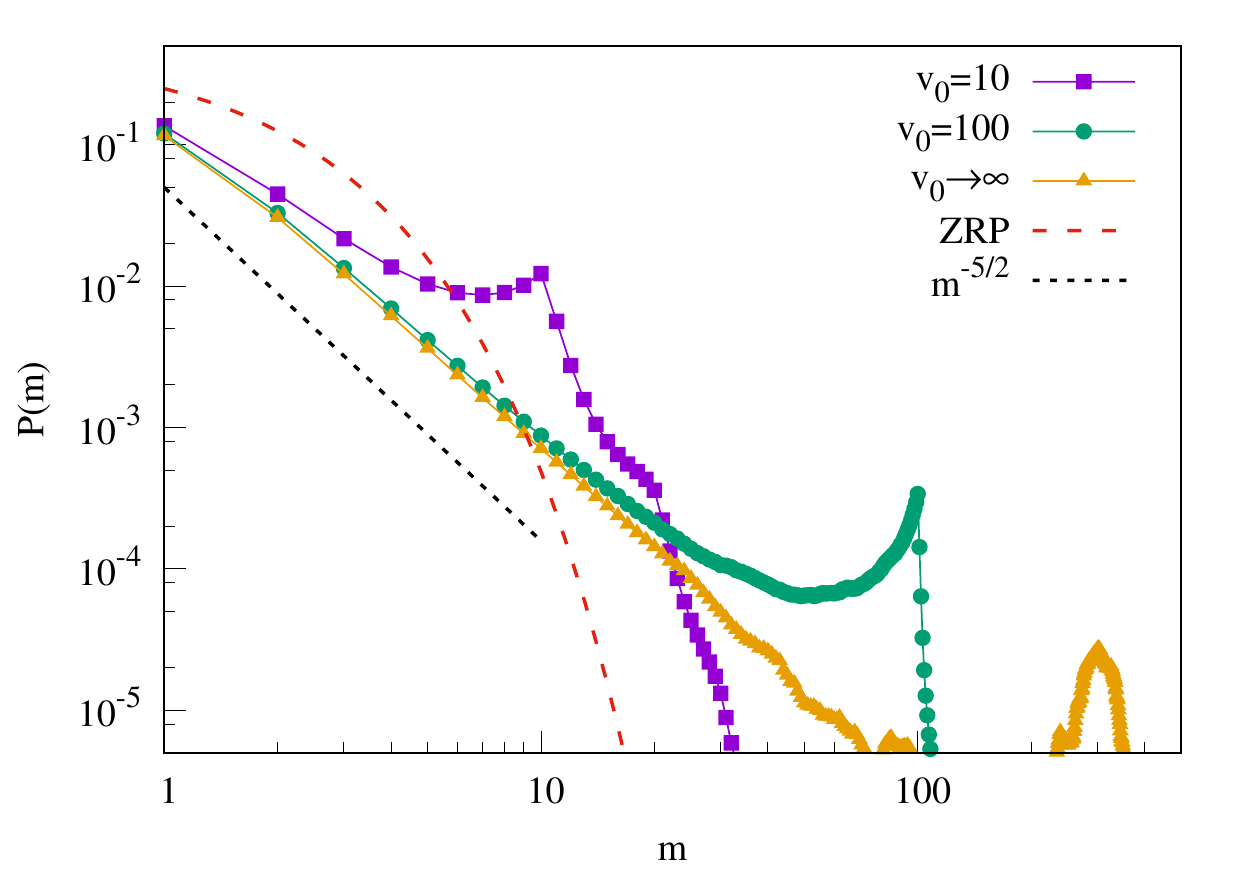}
\caption{ {\it Mass distributions in variant I}: Steady-state single-site mass distributions $P(m)$ are plotted as a function of mass $m$ for $v_0=$ 10 (magenta square), 100 (green circle), and $v_0 \rightarrow \infty$ (yellow triangle). The global density is kept fixed at $\rho=1$. For finite $v_0$, the distribution has peaks at $m$ equals to integer multiple of $v_0$.  As $v_0 \rightarrow \infty$, a macroscopic mass-condensate, along with a coexisting $m^{-5/2}$ power-law distributed fluid phase, is observed for $\rho > \rho_c=\sqrt{2}-1$.}
\label{ULHM_mass_distribution}
\end{center}
\end{figure}

To calculate the transport coefficients, we have to use the biased dynamics in Eq.~(\ref{General_Update_modify}) with $\phi(v)=\delta_{v,v_0}$ and also the indicator function $a_i^{v_0} =0$ (which is the case  in the limit of $v_0 \rightarrow \infty$). Subsequently, by putting $a_i^{v_0}=0$ in Eqs.~(\ref{local_g}) and (\ref{local_u}), we obtain the local observables $g(\rho)$ and $u(\rho)$,
\begin{eqnarray}
\label{local_g_Barma}
g(\rho) &=& \frac{1}{2}\left[ \rho + a(\rho) \right], \\
\label{local_u_Barma}
u(\rho) &=& \frac{1}{2}\left[ a(\rho) + \theta_2(\rho) \right],
\end{eqnarray} 
  as a function of density $\rho$. Then we substitute Eqs.~(\ref{occupancy_Barma}) and (\ref{third_moment_Barma}) into Eqs.~(\ref{local_g_Barma}) and (\ref{local_u_Barma}) and  use  Eq.~(\ref{transportKchip}), to find the bulk-diffusion coefficient $D(\rho)$ and the conductivity $\chi(\rho)$,
\begin{eqnarray}
\label{diffusivityBarma}
D(\rho) &=& \frac{1}{2} \frac{\partial g(\rho)}{\partial \rho} = \frac{1}{2(1+\rho)^2}, \\
\label{conductivityBarma}
\chi (\rho) &=& \frac{u(\rho)}{2} = \frac{\rho (1+\rho^2)}{2(1+\rho)(1-2\rho-\rho^2)},
\end{eqnarray}
respectively, as a function of density. Interestingly, the conductivity $\chi(\rho)$ as given in Eq.~(\ref{conductivityBarma}) diverges, or equivalently the resistivity (the inverse of the conductivity) vanishes,
at a critical density $\rho_c=\sqrt{2}-1$, exactly where the mass fluctuation also diverges [according to  Eq.~(\ref{fluctuation_Barma})]. Above the critical density, a macroscopic mass condensate forms in the system and coexists with the bulk phase with vanishing resistivity. Thus the clustering properties in this nonequilibrium mass aggregation model can be directly associated with the enhancement of the conductivity; in other words, the diverging conductivity is a dynamical manifestation of the underlying condensation transition and the diverging mass fluctuations in the system. Indeed, the intimate connection between transport and fluctuations is precisely encoded in the Einstein relation as following. The ratio of conductivity $\chi(\rho)$ and bulk-diffusion coefficient $D(\rho)$ from Eqs.~(\ref{conductivityBarma}) and (\ref{diffusivityBarma}) respectively, is given by
\be
\label{ER_Barma}
\frac{\chi(\rho)}{D(\rho)} = \frac{\rho (1+\rho)(1+\rho^2)}{1-2\rho-\rho^2} = \sigma^2(\rho),
\ee
which is nothing but the scaled variance given in Eq.~(\ref{fluctuation_Barma}) and immediately leads to the Einstein relation. Next we verify in simulations the Einstein relation Eq. ~(\ref{ER_Barma}). In Fig. \ref{FigHydroBM}, the scaled variance $\sigma^2(\rho)$ of subsystem mass obtained from simulations (circles) is plotted as a function of density and  compared to the ratio $\chi(\rho)/D(\rho)$ (line) obtained from the expressions in Eqs.~(\ref{diffusivityBarma}) and (\ref{conductivityBarma}); one could see the theory and simulations being in a reasonably good agreement. 
The condensate formation and the diverging mass fluctuation are also reflected in the corresponding single-site mass distribution plotted in Fig. \ref{ULHM_mass_distribution} for the global density $\rho > \rho_c$ (yellow triangles), where a delta peak along with a $m^{-5/2}$ power-law mass distribution is observed. Note that, while the conductivity diverges as we approach the transition point (from below), the bulk-diffusion coefficient remains finite. This implies that the phase transition in this case is facilitated not by vanishing diffusivity, but rather by a huge enhancement in the mobility of masses, which was also observed in mass transport processes studied in the context of  self-propelled particles \cite{Tanmoy_longhop}. 
However, in the presence of an infinite-density mass condensate in the coexistence phase (for global density $\rho > \rho_c$), where the condensate and the bulk fluid coexist with each other, the small-gradient expansions as in Eqs.~(\ref{grad-exp2}) and (\ref{grad-exp3}) break down and the calculations of the transport coefficients are no more valid. The break-down of  the hydrodynamic regime is reflected in that, for $\rho > \rho_c$, both the conductivity and the scaled mass fluctuation as in Eqs.~(\ref{conductivityBarma}) and (\ref{fluctuation_Barma}), respectively, become negative and clearly are not physical. This is  expected in the coexistence region, which is analogous to a first-order transition point in equilibrium, where hydrodynamic quantities are not defined exactly at the transition point; however, the transport coefficients calculated here are well defined if one approaches the transition point from below.

Following the approach  in Sec. \ref{Sec_Spreadingv2}, we now verify the expression of the bulk-diffusion coefficient  $D(\rho)$ by studying the density relaxation process, which is governed by the non-linear diffusion equation (\ref{hydroKchip5}) with the bulk-diffusion coefficient $D(\rho)$ given in Eq.~(\ref{diffusivityBarma}). We numerically integrate Eq.~(\ref{hydroKchip5}) by discretizing $x$ and $\tau$ and then using the Euler integration method. We take the initial density profile as given in Eq.~(\ref{initialKchip}) with the uniform background density  $\rho_0=0.1$, the strength of the perturbation $n_1=0.05$ and the width of the density perturbation $\Delta^2=5\times 10^{-3}$.
The results for the time evolution of the initial density profile is shown in the inset of Fig. \ref{FigHydroBM} for different times $\tau= 0$ (magenta circle), $2\times 10^{-2}$ (yellow triangle), and $10^{-1}$ (red square), where lines are the hydrodynamic theory and points are the simulation results; theory and simulations are in a quite good agreement over a decade of density values.

\subsection{Case IV: Other intermediate values of $v_0$}
\label{Sec_caseIV}

In this section, we numerically calculate the bulk-diffusion coefficient and the conductivity as a function of density for any finite values of $v_0$. In the previous Secs. \ref{Sec_ZRP}, \ref{Sec_v2}, and \ref{Sec_v_inf}, we have  calculated the transport coefficients for $v_0=1$, $2$ and $\infty$. However, for the intermediate values of $v_0$, calculating the local observables $g(\rho)$ and $u(\rho)$, which are required to calculate the transport coefficients, is more complicated even within the mean-field theory and presently beyond scope of this work. Therefore we now proceed with a numerical scheme to characterize the transport coefficients in these cases.

\begin{figure}
\begin{center}
\leavevmode
\includegraphics[width=9cm,angle=0]{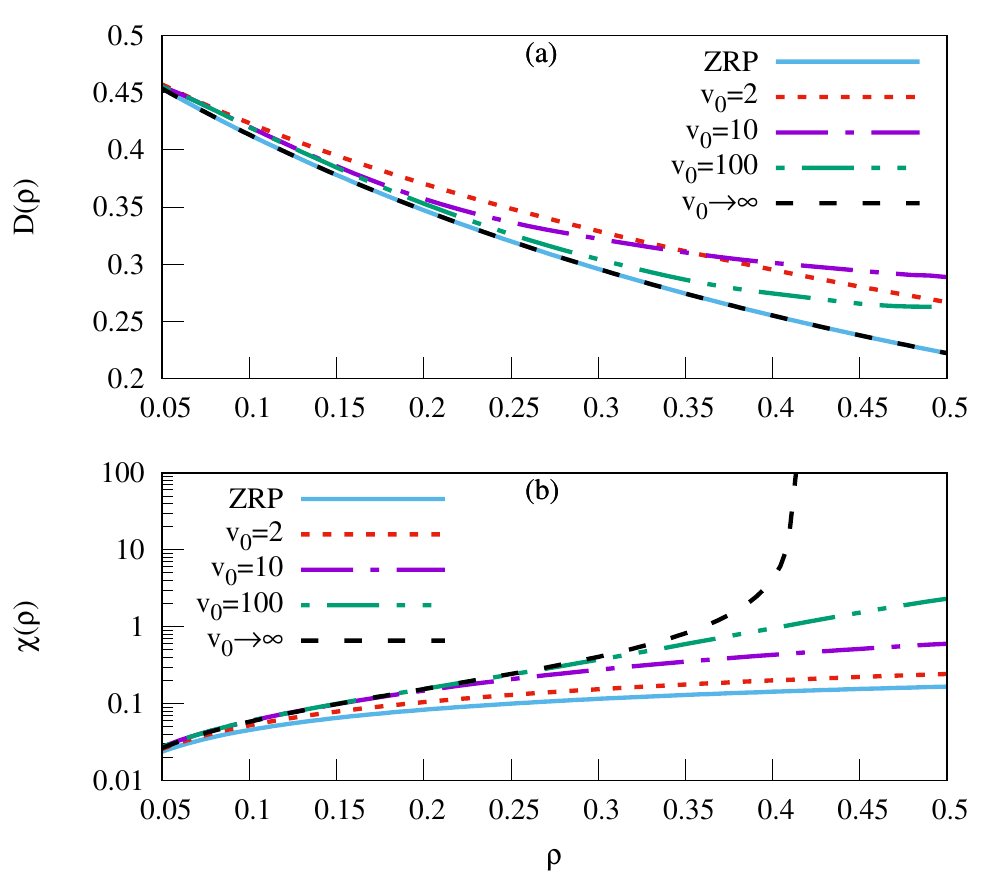}
\caption{ {\it Transport coefficients in variant I}: The bulk-diffusion coefficient $D(\rho)$ [panel (a)] and the conductivity $\chi(\rho)$ [panel (b)] are plotted as a function of density $\rho$ for $v_0=1$ [Eqs.~(\ref{ZRP_diff2}) and (\ref{ZRP_cond2})], 2 [Eqs. ~(\ref{D_v2}) and (\ref{Chi_v2})], 10 and 100 [numerically calculated using Eq.~(\ref{transportKchip})], and $v_0 \rightarrow \infty$ [Eqs.~(\ref{diffusivityBarma}) and (\ref{conductivityBarma})]. Interestingly, in the limit of $v_0 \rightarrow \infty$,  while the diffusion coefficient $D(\rho)$ remains finite, the conductivity $\chi(\rho)$ diverges at the critical density $\rho_c=\sqrt{2}-1$, signifying a condensation transition in the system.}
\label{ULHMtransport}
\end{center}
\end{figure}

To obtain the quantities $g(\rho)$ and $u(\rho)$, as given in Eqs. ~(\ref{local_g}) and (\ref{local_u}), one first needs to calculate the steady-state mass distributions. By performing Monte Carlo simulations, we compute the steady-state single-site mass distribution $P(m)$, which shows some interesting features. In Fig. \ref{ULHM_mass_distribution}, we plot probability distribution $P(m)$ as a function of mass $m$ at a single site for $v_0=10$ (magenta squares), $100$ (green circles) and for $v_0 \rightarrow \infty$ (yellow triangles) keeping global density $\rho=1$. The distributions are compared with that for ZRP in Eq.~(\ref{ZRP_dist}) (red dashed line). For a finite $v_0$, we notice that the distributions have peaks at mass values being equal to integer multiple of $v_0$. However the most striking observation here is the formation of a macroscopic size mass-condensate in the limit of of $v_0 \rightarrow \infty$ and beyond a critical density $\rho_c = \sqrt{2}-1$ (we have taken $p=q=1/2$). In the translation-symmetry broken condensate phase, the excess mass of amount $L(\rho-\rho_c)$ coexists with a bulk  phase, having $m^{-5/2}$ power-law single-site mass distribution and diverging conductivity (i.e., vanishing resistivity). We use the mass distribution $P(m)$ in Eqs. ~(\ref{local_g}) and (\ref{local_u}) to obtain the quantities $g(\rho)$ and $u(\rho)$ and thus to calculate the bulk diffusion coefficient $D(\rho)$ and conductivity $\chi(\rho)$ using Eq.~(\ref{transportKchip}). 
In panel (a) of Fig. \ref{ULHMtransport}, we plot numerically calculated  $D(\rho)$ as a function of  density $\rho$ for $v_0=10$ (magenta dashed-dotted line) and $100$ (green dash-dot-dot line) along with those observed analytically for $v_0=1$ (ZRP, blue solid line), $2$ (red dotted line) and $v_0 \rightarrow \infty$ (black dashed line) given by Eqs.~(\ref{ZRP_diff2}), (\ref{D_v2}) and (\ref{diffusivityBarma}), respectively. Interestingly, the bulk-diffusion coefficient $D(\rho)$, though  a monotonically decreasing function of density, does not vanish and remains finite even on the onset of cluster formation in the system. This is unlike the clustering observed near an equilibrium critical point, where the bulk-diffusion coefficient vanishes. Moreover, we see that $D(\rho)$ for $v_0=1$ and $\infty$ are the same as given by Eqs.~(\ref{ZRP_diff2}) and (\ref{diffusivityBarma}), respectively. This implies that the bulk diffusion coefficient must be a non-monotonic function of $v_0$.
Similarly, we also plot numerically calculated conductivity $\chi(\rho)$ as a function of density $\rho$ in panel (b) of Fig. \ref{ULHMtransport} for $v_0=10$ (magenta dash-dot line) and $100$ (green dash-dot-dot line) along with those obtained analytically for $v_0=1$ (ZRP, blue solid line), $2$ (red dotted line) and $v_0 \rightarrow \infty$ (black dashed line) given by Eqs.~(\ref{ZRP_cond2}), (\ref{Chi_v2}) and (\ref{conductivityBarma}), respectively. Clearly, unlike the diffusivity, the conductivity $\chi(\rho)$ increases monotonically with increasing $v_0$, and diverges at a critical density $\rho=\rho_c=\sqrt{2}-1$ when $v_0 \rightarrow \infty$. This is a clear indication of a mobility-driven clustering in the system. That is, from the dynamical point of view, the increasing mobility of masses drives the clustering, thus resulting in large mass fluctuations and, beyond a critical density, leading to the condensation transition in the system.

\subsubsection{Density relaxation and verification of the Einstein relation}
\label{SecER_ULHM}

\begin{figure}[h]
\begin{center}
\leavevmode
\includegraphics[width=9cm,angle=0]{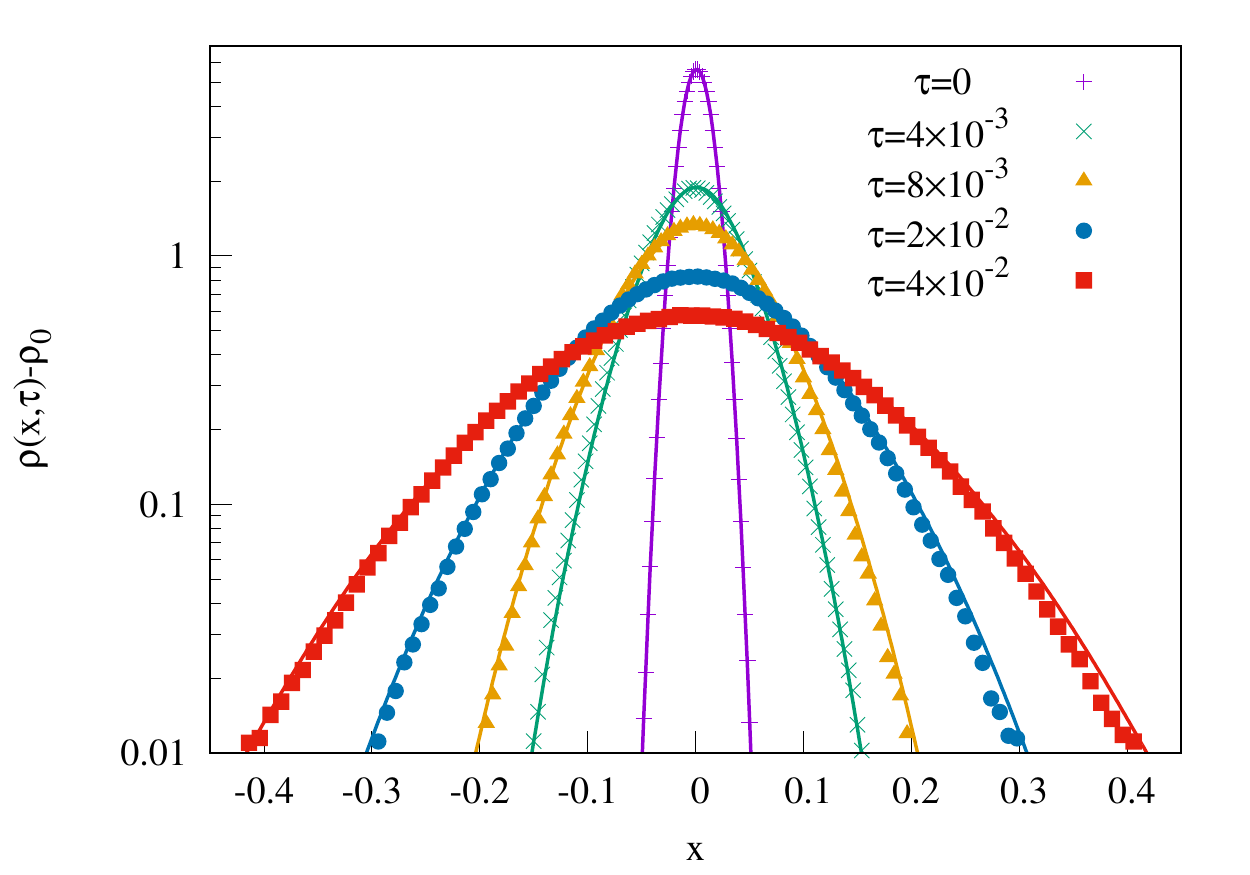}
\caption{ {\it Density relaxation in variant I, $v_0 =10$:} We verify the functional dependence of the bulk-diffusion coefficient $D(\rho)$ on density $\rho$ through density relaxations at various hydrodynamic times $\tau=0$ (magenta plus), $4\times 10^{-3}$ (green cross), $8\times 10^{-3}$ (yellow triangle),  $2\times 10^{-2}$ (blue circle), and $4\times 10^{-2}$ (red square); we take initial condition Eq.~(\ref{initialKchip}) with $\rho_0=0.5$, $n_1=0.2$ and $\Delta^2=2\times10^{-4}$. The lines denote the numerically integrated hydrodynamic time-evolution  and points denote simulation results.}
\label{FigHydroULHM}
\end{center}
\end{figure}

Now we verify the theoretical results for the bulk-diffusion coefficient $D(\rho)$ by studying  density relaxation for which we follow the procedure employed in Sec. \ref{Sec_Spreadingv2}. The only difference in this case is that now we do not have an explicit analytic expression for the diffusivity $D(\rho)$, rather we have only the numerical values of the diffusivity at different densities. This however help us to numerically integrate Eq.~(\ref{hydroKchip5}), starting from the initial distribution given by Eq.~(\ref{initialKchip}). We set $v_0=10$, $\rho_0=0.5$, $n_1=0.2$, and $\Delta^2=2\times10^{-4}$.  In Fig. \ref{FigHydroULHM}, we plot the density profiles obtained from hydrodynamics (lines) and direct simulations (points) at different times $\tau=0$ (magenta plus), $4\times 10^{-3}$ (green cross), $8\times 10^{-3}$ (yellow triangle),  $2\times 10^{-2}$ (blue circle), and $4\times 10^{-2}$ (red square). We again see that the diffusivity obtained using the mean field theory captures the simulation results quite well.

\begin{figure}[h]
\begin{center}
\leavevmode
\includegraphics[width=9cm,angle=0]{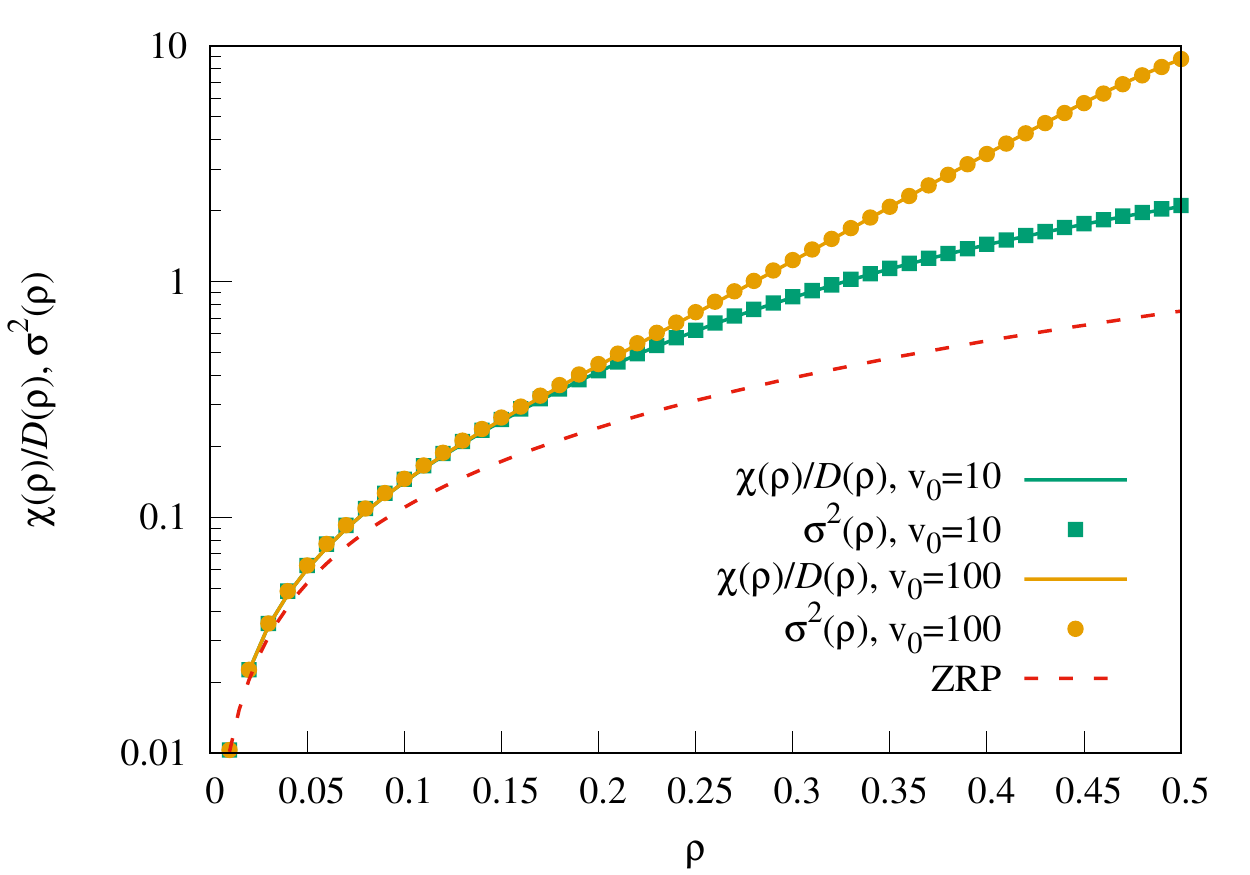}
\caption{ {\it Verification of the Einstein relation in variant I:} Scaled subsystem mass fluctuation $\sigma^2(\rho)$ is plotted as a function of density $\rho$ for $v_0=10$ (green square) and 100 (yellow circle). It is compared with the ratio of two transport coefficients $\chi(\rho)$ and $D(\rho)$, respectively, calculated numerically using Eq.~(\ref{transportKchip}), for $v_0=10$ (green solid line) and $100$ (yellow solid line). Simulations (points) and hydrodynamic theory (lines) agree quite well, implying the existence of Einstein relation in the system. For comparison, the scaled subsystem mass fluctuation for the ZRP is also plotted (red dashed line).}
\label{ULHM_ER}
\end{center}
\end{figure}

To verify the Einstein relation, we plot  in Fig. \ref{ULHM_ER} the ratio of the two transport coefficients $\chi(\rho)/D(\rho)$ as a function of density $\rho$ for $v_0=10$ (green solid line) and $100$ (yellow solid line). Subsequently, we compute from simulation the scaled steady-state mass fluctuation $\sigma^2(\rho)$ and plot in Fig. \ref{ULHM_ER} the scaled fluctuation $\sigma^2(\rho)$ as a function of density for $v_0=10$ (green square) and $100$ (yellow circle) along with that for ZRP as in Eq.~(\ref{ZRP_fluctuation}) (red dashed line). A nice agreement between points and lines demonstrates the existence of the Einstein relation also for finite values of  $v_0$.

Note in Fig. \ref{ULHM_mass_distribution} that, with increasing $v_0$,  the mass distributions develop greater weights at their tails. This behavior points to large mass fluctuations due to the formation of larger clusters of masses for higher values of $v_0$ -  the fact which is also reflected in Fig \ref{ULHM_ER}. Finally, in the limit of $v_0 \rightarrow \infty$, the system goes through a condensation transition and a single macroscopic cluster of mass is formed. Indeed, as demonstrated in Fig. \ref{ULHMtransport}, the conductivity $\chi(\rho)$ is solely responsible for the mass clustering processes as, on the approach towards the clustering, it increases monotonically with increasing $v_0$, whereas the bulk diffusivity $D(\rho)$ remains finite. 
Clearly, the clustering phenomena in these mass aggregation processes cannot be associated with the vanishing diffusivity, rather they are driven by the large conductivity, which implies a mobility-driven clustering. 
In the next section, we consider another variant with the choice of exponential distribution $\phi(v) \propto \exp(-v/v_*)$, which is motivated by one dimensional run-and-tumble-particles dynamics \cite{Soto_PRE}. We explore whether the above mentioned scenario of mobility-driven clustering extends also to this variant.

\section{Variant II: Fragmentation of random amount of mass}
\label{Sec_Var2}

In this section, we consider another variant of the mass aggregation process by choosing the distribution of random variable $v$  to be  exponentially distributed,
\begin{equation}
\phi(v)=(1-e^{-1/v_*}) e^{-v/v_*},
\label{phi}
\end{equation}
with $v_*$ being the cut-off to the distribution $\phi(v)$. Depending on the presence (or the absence) of the single-particle chipping, we consider the following two cases separately: (i) $p=0$ and $q=1$ (no chipping and only fragmentation), and (ii) $p=1/2$ and $q=1/2$ (both chipping and fragmentation). Below we calculate the transport coefficients and demonstrate the Einstein relation for these models. Indeed, in both the cases, the transport coefficients can be calculated analytically in the limit of $v_* \rightarrow \infty$.

\subsection{Case I: $p=0$ and $q=1$}
\label{Sec_Var2_Case1}

Here we consider the case in the absence of the single-particle chipping and choose $p=0$ and $q=1$.  We first argue that, in the limit of $v_* \gg 1$, the system initially undergoes an aggregation process. In the steady-state, mass is concentrated on a few sites with large clusters on them, with most of the other sites being empty. We will also present the numerical findings for finite $v_*$ later in this section.

Let us start with a uniform distribution of particles on the lattice. Initially, almost all moves involve a complete transfer of the mass onto a neighboring site, since the threshold $v$ is usually greater than the mass on the site. Denoting the sites with mass on them as ``$A$", the process has the form $A + A \rightarrow A$ with diffusing ``$A$" type masses \cite{Krapivsky-review, burschka1989transition, ben1990statics}.
This process stops when the number of empty sites, and the average mass per site, become large enough that mass transfer events start increasing the number of occupied sites. In other words, we can say that the reverse process $A \rightarrow A + A$ starts becoming significant. In the steady state, the two processes balance each other and we have a stationary mass distribution on a site. We assume that the mass on occupied sites is distributed according to an exponential distribution, where the average mass on a site is large. Consequently the probability distribution  $P_{occ}(m)$ of mass at a site, provided the site is occupied, is given by
\be
P_{occ}(m) = (e^{\eta} - 1) e^{-\eta m} ,
\label{eq:Pocc}
\ee
with the parameter $\eta$ being determined by the mean mass of an occupied site,
\be
\left\langle m \right\rangle  = (e^{\eta}-1)^{-1} .
\label{eq:avgm}
\ee
We now denote $r$ as the probability that a mass transfer process does not split the mass at a site. This is simply the probability that $m < v$, where $m$ is the mass on the site and $v$ is the fragmentation threshold, chosen according to Eq.~(\ref{phi}). Then $r$ is given by
\be
r \equiv \sum_{m=2} \frac{1}{v_*} e^{-m/v_*} (e^{\eta}-1) e^{-\eta m} + (1-e^{-\eta}).
\ee
The last term is due to the fact that if the site initially has $m=1$, this cannot be split into two. Since the mass transfer events themselves happen at unit rate per site, the rates for the diffusion and fragmentation processes are ($A$ denotes a site with mass and $\phi$ an empty site)
\bea \nonumber
A \phi \xrightleftharpoons [r]{r} \phi A,
\\ \nonumber
A \phi \xrightarrow {(1-r)} A A,
\\ \nonumber
\phi A \xrightarrow {(1-r)} A A.
\eea

\begin{figure}[h]
\begin{center}
\leavevmode
\includegraphics[width=9cm,angle=0]{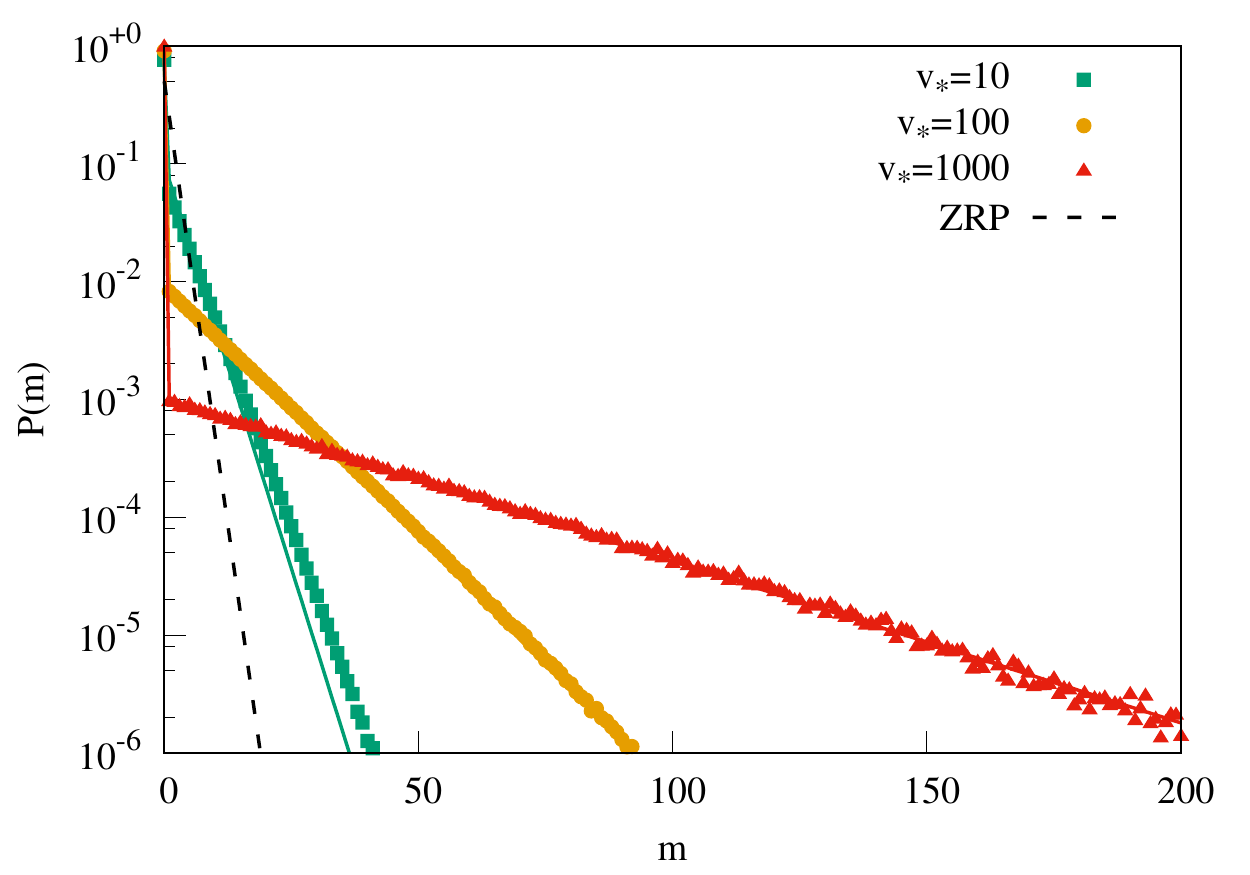}
\caption{ {\it Mass distributions in variant II, $p=0$, $q=1$:} Single-site mass distributions $P(m)$ are plotted as a function of mass $m$ for $v_*=10$ (green square), $100$ (yellow circle), and $1000$ (red triangle); lines denote the corresponding distributions obtained analytically [Eq.~(\ref{dist_p0_q1})] for $v_*=10$ (green line), $100$ (yellow line), and $1000$ (red line). The mass distributions are compared with that for ZRP (black dashed line) as in Eq.~(\ref{ZRP_dist}).}
\label{gULHM_mass_distribution_p0}
\end{center}
\end{figure}
We now use the empty interval method \cite{burschka1989transition, ben1990statics} to find out the stationary distribution of the mass and empty-site clusters. Let $E_n$ denotes the probability that a chosen segment of $n$ contiguous sites is empty. The equation for the empty intervals due to the above processes is
\be
\frac{d E_n}{dt} = 2 r (E_{n+1} + E_{n-1}- 2 E_n) + 2 (1-r) (E_n - E_{n+1}).
\ee
The solution of this equation is given by
\be
E_n = c r^n,
\ee
and the probability distribution of void sizes can be derived from $E_n$ through the relation,
\be
P_n = E_{n+2} + E_{n} - E_{n+1} = c (1-r)^2 r^n.
\ee
After normalization, we determine that $c = (1-r)^{-1}$. Thus the average void size is
\be
\left\langle n \right\rangle = \frac{r}{1-r}, \label{eq:avgn}
\ee
and, provided the average density is $\rho$, we have the relation
\be
\rho (\left\langle n \right\rangle +1) = \left\langle m \right\rangle.
\ee
Using Eqs.~(\ref{eq:avgm}) and (\ref{eq:avgn}), and solving for $\eta$ in terms of $v_*$ and $\rho$, we get
\be
\eta = \mu(\rho) = \sqrt{\frac{1}{v_* \rho}} +  {\cal O}\left(v_*^{-1}\right).
\ee
The number of occupied sites is therefore given by
\be
n_c = \sqrt{\frac{\rho}{v_*}} + O\left(v_*^{-1}\right).
\ee
Assuming that sites are independently occupied with probability $n_c$, we have the mass distribution
\be
\label{dist_p0_q1}
P(m) = 
\begin{cases}
	1-n_c= 1 - \sqrt{\frac{\rho}{v_*}} & \mbox{ for $m=0$,}\\
	n_c P_{occ}(m)= \frac{1}{v_*} ~ e^{-\sqrt{\frac{1}{\rho v_*}} m} &\mbox{ for $m>0$.}
\end{cases}
\ee
Similar mass distributions have been observed in a system related to hardcore run-and-tumble particles on a one dimensional lattice \cite{Rahul-Subhadip-Rajesh}.

\begin{figure}[h]
\begin{center}
\leavevmode
\includegraphics[width=9cm,angle=0]{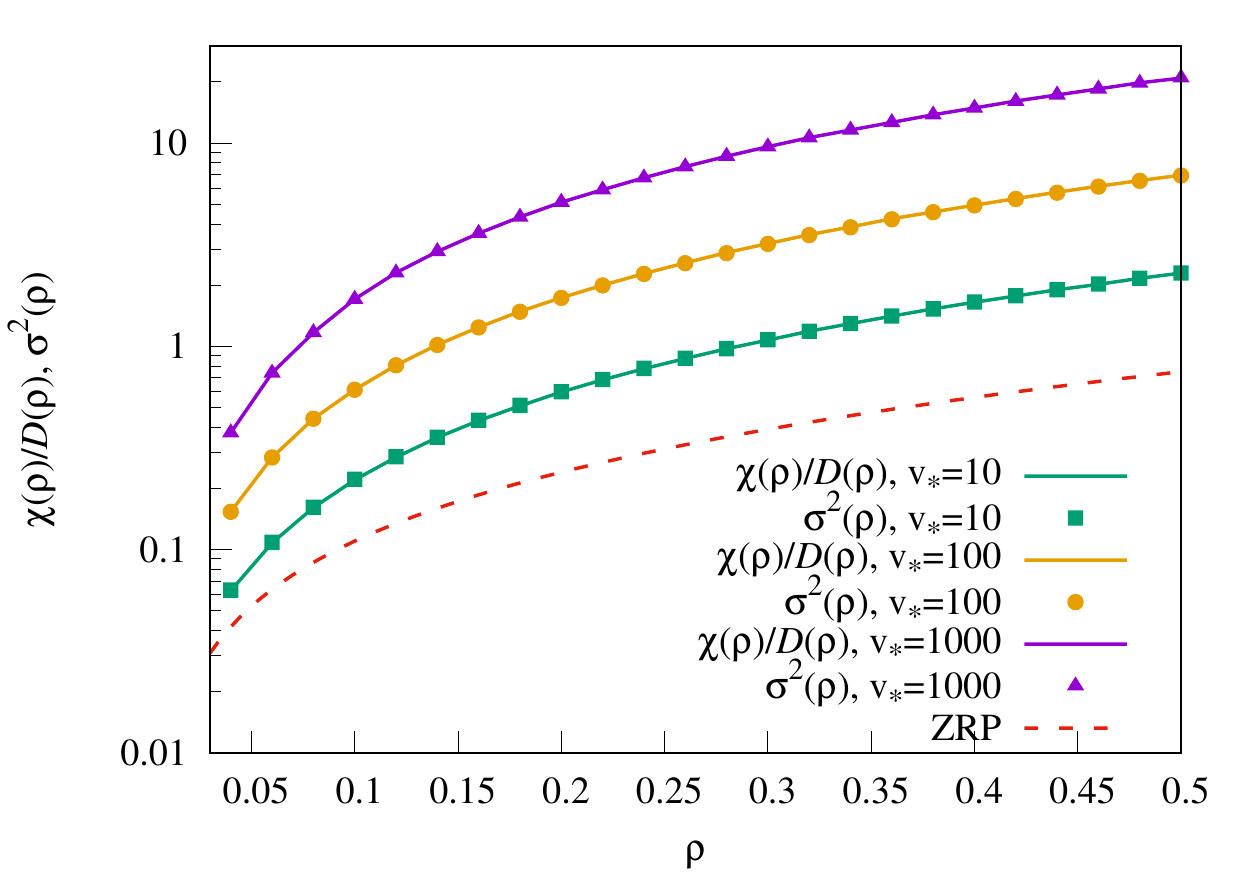}
\caption{ {\it Verification of the Einstein relation in variant II, $p=0$ and $q=1$:} Simulation results for the scaled subsystem mass fluctuation $\sigma^2(\rho)$ is plotted as a function of density $\rho$ for $v_*=10$ (green square), 100 (yellow circle) and 1000 (magenta triangle). It is compared with the ratio of two transport coefficients $\chi(\rho)$ and $D(\rho)$, respectively, calculated numerically using Eq.~(\ref{transportKchip}), for $v_*=10$ (green solid line), $100$ (yellow solid line) and $1000$ (magenta solid line). Simulations (points) and hydrodynamic theory (lines) agree quite well, thus demonstrating the existence of the Einstein relation in the system. For comparison, the scaled subsystem mass fluctuation $\sigma^2(\rho)$ for the ZRP (red dashed line) is also plotted.}
\label{UgLHM_ER_p0}
\end{center}
\end{figure}

In Fig. \ref{gULHM_mass_distribution_p0}, we plot the analytically calculated single-site mass distribution $P(m)$ with mass $m$ given by Eq.~(\ref{dist_p0_q1}) for $v_*=10$ (green line), $100$ (yellow line), and $1000$ (red line), and compare them with the same calculated from Monte Carlo simulation for $v_*=10$ (green square), $100$ (yellow circle), and $1000$ (red triangle). The results are also compared with the same for the ZRP (black dashed line) given by Eq.~(\ref{ZRP_dist}). The plots show an excellent agreement between the mean-field theory and the simulation results in the limit of $v_*$ large. One can now use the expression of single-site mass distribution $P(m)$ in Eq.~(\ref{dist_p0_q1}) to calculate the scaled mass fluctuations in the large $v_*$ limit as
\be
\sigma^2(\rho) = 2\sqrt{v_*\rho^3} + {\cal O}(v_*^{-1}),
\ee
where we have assumed that the nearest correlations vanish in the limit of system size large (see Fig. \ref{Fig:Scaled_correlation_var2} in Appendix \ref{app_3}). Now, by putting $p=0$ and $q=1$ in Eqs.~\eqref{g_i_general} and \eqref{u_i_general}, we can write the local observables as 
	\begin{align}
	g(\rho) =   \left[1- \phi(0)-\phi(1)\right] \rho - \sum_{v=2}^{\infty} \phi(v) \sum_{m=v}^{\infty} m P(m) \nonumber \\ + \sum_{v=1}^{\infty} \phi(v) v \sum_{m=v}^{\infty} P(m) , \label{eq:g}\\
	u(\rho) = \left[1- \phi(0)-\phi(1)\right]\la m^2 \ra + \sum_{v=1}^{\infty} \phi(v) v^2 \sum_{m=v}^{\infty} P(m) \nonumber \\ - \sum_{v=2}^{\infty} \phi(v) \sum_{m=v}^{\infty} m^2 P(m), \label{eq:u}
	\end{align}
which, along with Eq.~(\ref{transportKchip}), immediately lead to the desired transport coefficients. The explicit expressions of the transport coefficients are however quite complicated for any particular $v_*$, but they do have a simple asymptotic form as we show next.  
\begin{figure}[h]
\begin{center}
\leavevmode
\includegraphics[width=9cm,angle=0]{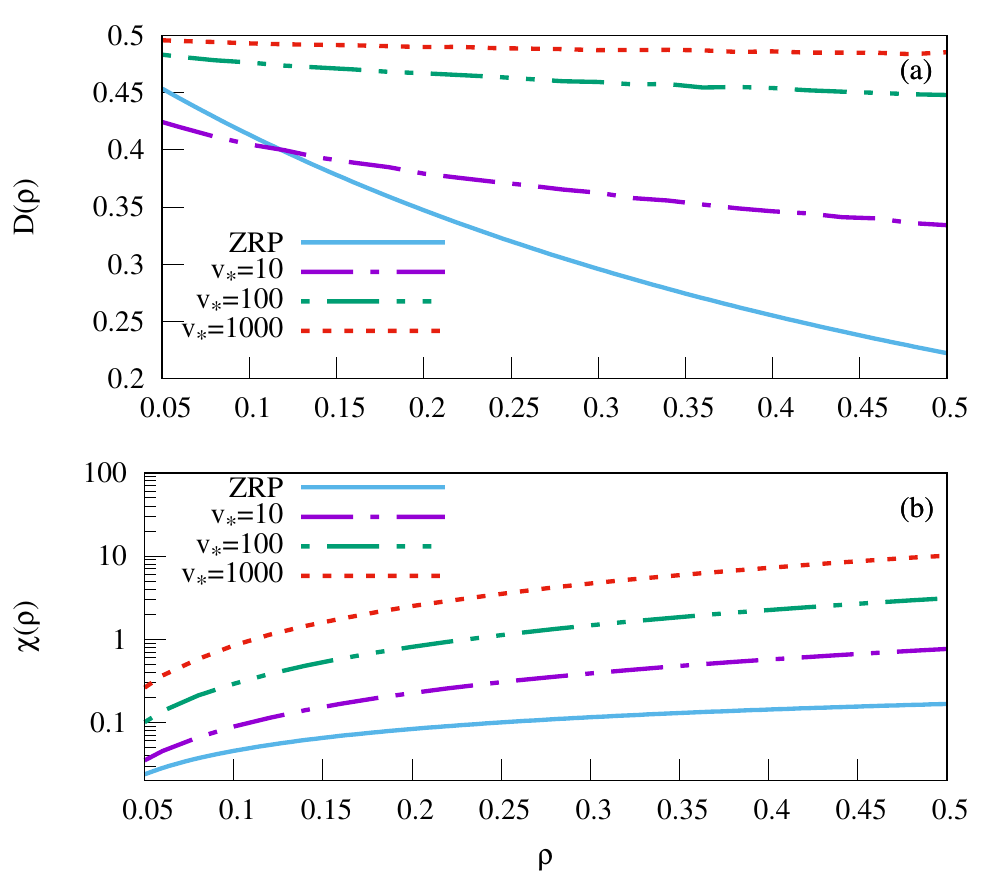}
\caption{ {\it Transport coefficients in variant II, $p=0,~q=1$:} The bulk-diffusion coefficient $D(\rho)$ [panel (a)] and the conductivity $\chi(\rho)$ [panel (b)] are plotted as a function of density $\rho$ for $v_*=10$, $100$ and $1000$. While the bulk-diffusion coefficient remains finite for all $v_*$, the conductivity however monotonically increases with increasing $v_*$. We compare the transport coefficients with that (sky blue solid lines) for the ZRP [Eqs.~(\ref{ZRP_diff2}) and (\ref{ZRP_cond2})].}
\label{UgLHMtransport_p0}
\end{center}
\end{figure}
In the limit of large $v_*$, we can explicitly determine the asymptotic functional form of the bulk-diffusion coefficient $D(\rho)$ and the conductivity $\chi(\rho)$ by using the expressions (\ref{eq:g}) and (\ref{eq:u}) and expanding the expressions in the leading order of $v_*$,
\bea \label{D_large_v*}
D(\rho) &=& \frac{1}{2} - \frac{3}{2} \sqrt{\frac{\rho}{v_*}},  
\\
\label{chi_large_v*}
\chi(\rho) &=& \sqrt{v_* \rho^3}.
\eea
in this limiting case, we also obtain the ratio of the transport coefficients, which can be related to the scaled mass fluctuation $\sigma^2(\rho)$ as 
\be
\frac{D(\rho)}{\chi(\rho)} = \sqrt{\frac{1}{4 v_*\rho^3}} +  {\cal O}(v_*^{-1}) = \frac{1}{\sigma^2(\rho)},
\ee
immediately implying the Einstein relation. We now verify in simulations the Einstein relation for finite values of $v_*$. To this end, we first calculate mass distribution $P(m)$ from simulation and use it in Eqs.~(\ref{eq:g}) and (\ref{eq:u}) to calculate the two local observables $g(\rho)$ and $u(\rho)$ as a function of density $\rho$. Then, by substituting $g(\rho)$ and $u(\rho)$ in Eq.~(\ref{transportKchip}), we obtain the bulk-diffusion coefficients $D(\rho)$ and the conductivity $\chi(\rho)$. In Fig. \ref{UgLHMtransport_p0}, we plot $D(\rho)$ (panel (a)) and $\chi(\rho)$ (panel (b)) as a function of $\rho$ for $v_*=10$ (magenta dash-dot), 100 (green dash-dot-dot) and 1000 (red dotted) and compared them with those obtained for ZRP (sky blue solid lines) given by Eqs.~(\ref{ZRP_diff2}) and (\ref{ZRP_cond2}). Now, to check the Einstein relation, in Fig. \ref{UgLHM_ER_p0} we plot the ratio of the two transport coefficients $\chi(\rho)/D(\rho)$ as a function of density $\rho$  for $v_*=10$ (green solid line), $100$ (yellow solid line) and $1000$ (magenta solid line) and compare them with the scaled mass fluctuations $\sigma^2(\rho)$ computed from simulation for $v_*=10$ (green square), 100 (yellow circle) and 1000 (magenta triangle). An excellent agreement between simulations (points) and the theory (lines) demonstrates the existence of the Einstein relation for finite $v_*$.

Furthermore, we note that the scaled subsystem mass fluctuation in this variant of the mass aggregation processes increase with  increasing $v_*$, as seen in the plots for the single-site mass distributions in Fig \ref{gULHM_mass_distribution_p0}, where the probability weights for larger masses grow with increasing $v_*$; for any nonzero finite density, the mass fluctuation diverges in the limit of $v_* \rightarrow \infty$. Moreover, in Fig.~\ref{UgLHMtransport_p0}, we find that, while the conductivity  increases with increasing $v_*$ in an unbounded fashion, the bulk-diffusion coefficient remains bounded, decreases with increasing $v_*$ and eventually saturates to a finite value at very large $v_*$. 
This clearly demonstrates the scenario of mobility-driven clustering in the system, where, as evident through the Einstein relation, the diverging conductivity (or, equivalently, the mobility) contributes to the diverging mass fluctuations. We discuss below another variant of mass aggregation model, where the chipping process is also present and the system exhibits a condensation transition in the limit of $v_* \rightarrow \infty$ analogous to the case III of the variant I.

\subsection{Case II: $p=q=1/2$}

In this variant, we include the single-particle chipping dynamics along with the fragmentation process with the exponentially distributed $v$. In this case, except in the limit of $v_* \rightarrow \infty$, it is difficult to obtain a closed form expression of the transport coefficients. Therefore, here we resort to the numerical scheme prescribed in Sec. \ref{Sec_caseIV} for all finite $v_*$.

\begin{figure}[h]
\begin{center}
\leavevmode
\includegraphics[width=9cm,angle=0]{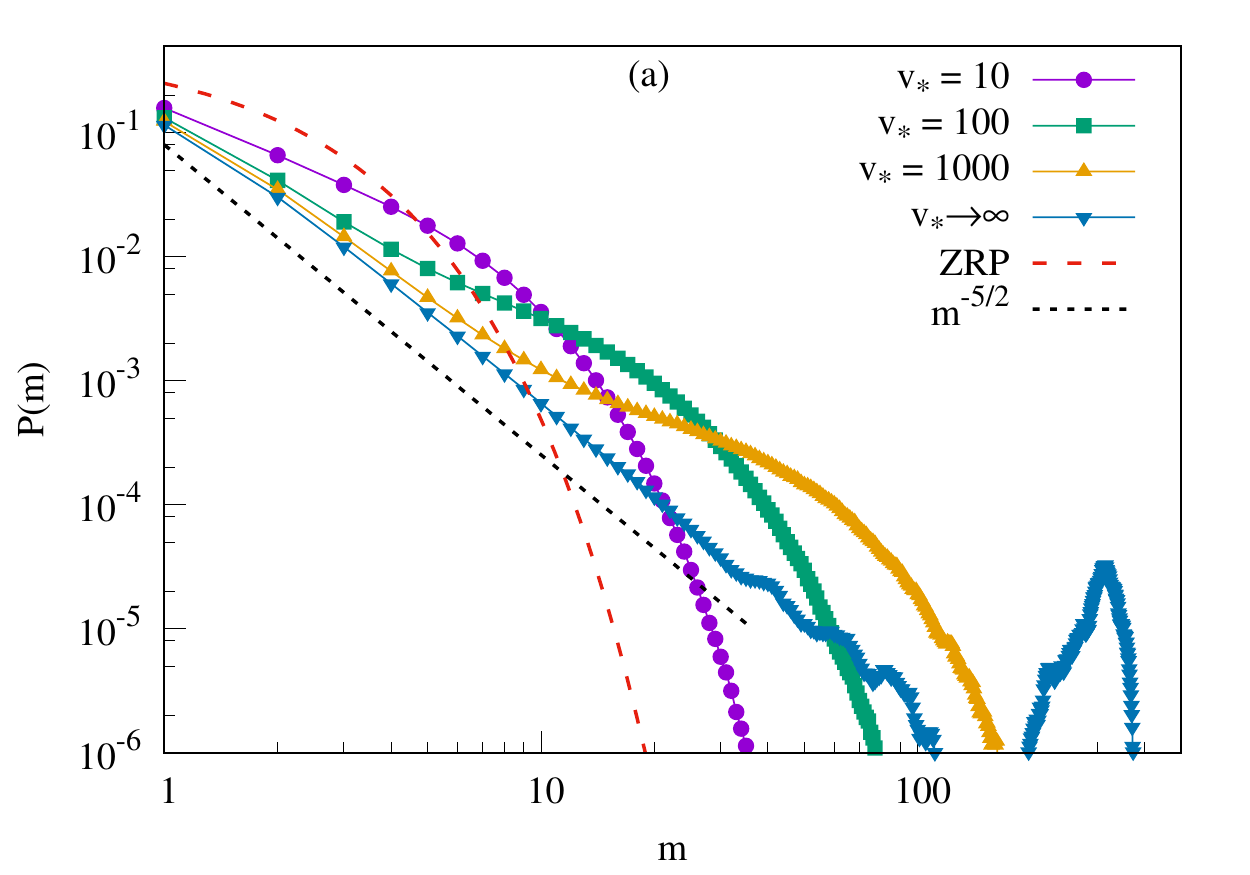}
\includegraphics[width=9cm,angle=0]{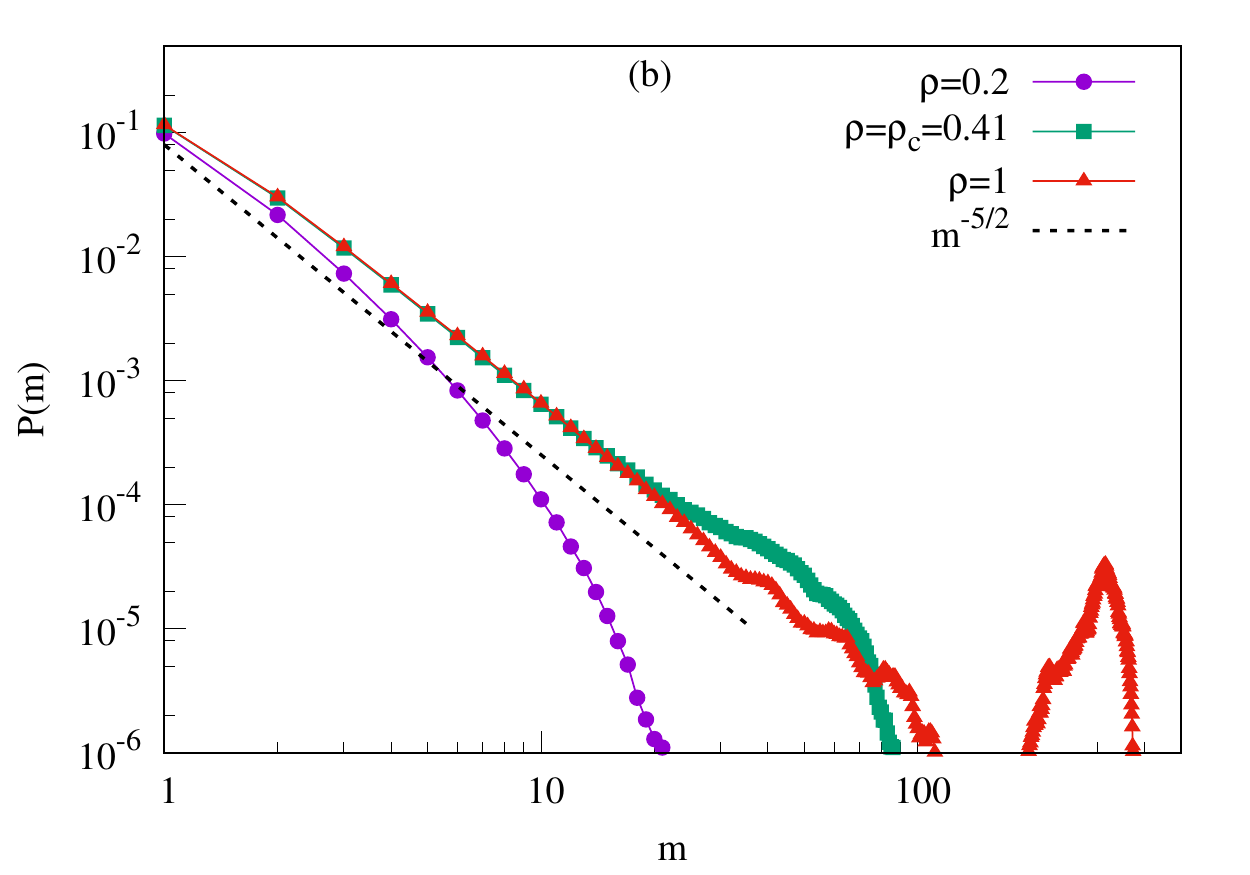}
\caption{ {\it Mass distributions in variant-II, $p=q=1/2$:} Steady-state single-site mass distributions $P(m)$ are plotted as a function of mass $m$. Panel (a): The distributions are plotted for $v_*=10$ (magenta circle), $100$ (green square), $1000$ (yellow triangle) and $v_* \rightarrow \infty$ (blue inverted triangle) for a fixed density $\rho=1$. Panel (b): The dependence of the distributions on density in the limit $v_*\rightarrow \infty$ is shown for various densities $\rho=0.2$ (magenta circle), $0.41$ (green square), and $1$ (red triangle).  For $v_* \rightarrow \infty$, a macroscopic mass-condensate, coexisting with a $m^{-5/2}$ power-law distributed fluid phase, is observed beyond a critical density $\rho_c$.}
\label{gULHM_mass_distribution}
\end{center}
\end{figure}

\begin{figure}[h]
\begin{center}
\leavevmode
\includegraphics[width=9cm,angle=0]{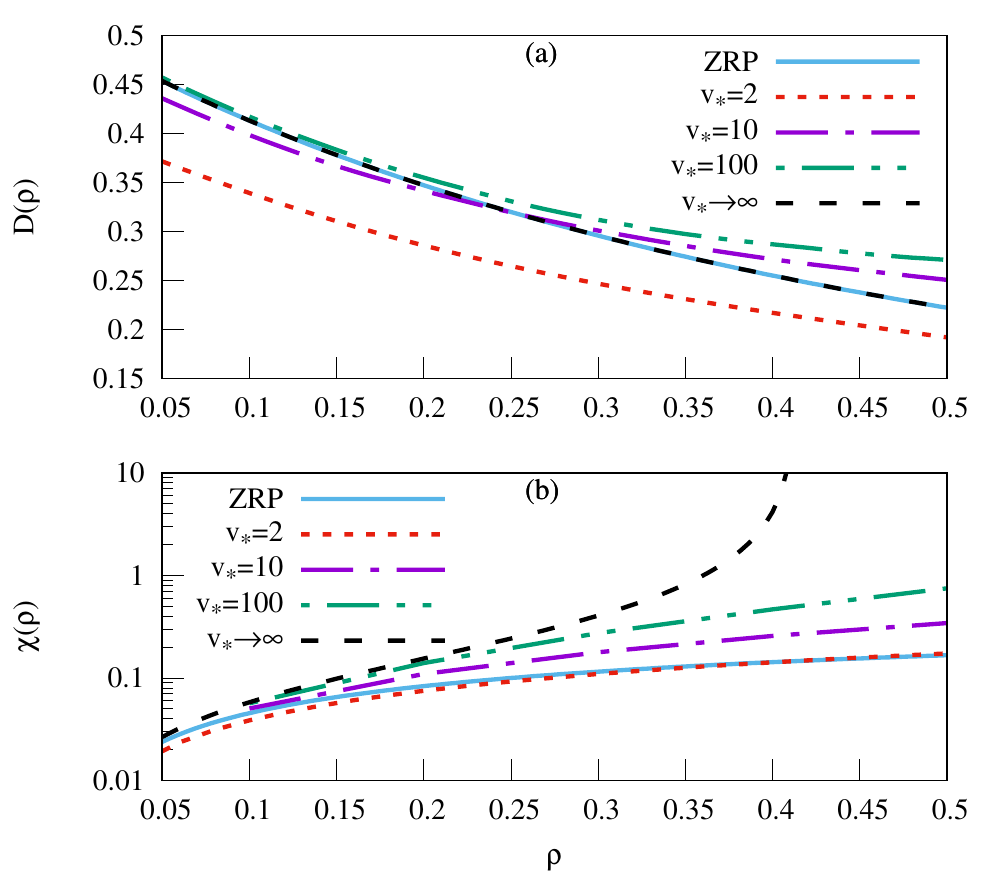}
\caption{ {\it Transport coefficients in variant II, $p=q=1/2$:} The bulk-diffusion coefficient $D(\rho)$ [panel (a)] and the conductivity $\chi(\rho)$ [panel (b]) are plotted as a function of density $\rho$ for $v_*=2$, $10$, $100$ and $\infty$; the two transport coefficients are calculated numerically using Eq.~(\ref{transportKchip}). In this variant too, while the bulk-diffusion coefficient remains finite, the conductivity increases with increasing $v_*$. We compare them with that for ZRP  (sky blue solid lines), as in Eqs.~(\ref{ZRP_diff2}) and (\ref{ZRP_cond2}). }
\label{UgLHMtransport}
\end{center}
\end{figure}

We begin by calculating the steady-state single-site mass distributions $P(m)$ from simulations. We plot the mass distributions as a function of mass $m$ in panel (a) of Fig. \ref{gULHM_mass_distribution} for $v_*=10$ (magenta circle), $100$ (green square), $1000$ (yellow triangle) and $v_* \rightarrow \infty$ (blue inverted triangle) for a fixed density $\rho=1$. For comparison, in the same figure, we also plot the mass distributions for the ZRP [Eq.~(\ref{ZRP_dist})] (red dashed line). In these cases too, for larger $v_*$,  we find that the mass distributions have a longer tails, i.e., whose weights increase with increasing $v_*$.  Finally, in the limit of $v_*\rightarrow \infty$, the dynamics becomes equivalent to the case of variant I with $v_0 \rightarrow \infty$ discussed in Sec. \ref{Sec_v_inf}. For $v_* \rightarrow \infty$, we have calculated the occupation probability $a(\rho)$ from simulations and compared $a(\rho)$ with that in Eq.~(\ref{occupancy_Barma}), which is in good agreement with the simulation results (not shown here). Therefore we observe that, in this case also, the competition between  chipping and aggregation (together with diffusion) results in a condensation transition beyond a critical density $\rho_c=\sqrt{2}-1$. In panel (b) of Fig. \ref{gULHM_mass_distribution}, we plot the mass distribution $P(m)$ as a function of $m$ for densities $\rho=0.2$ (magenta circle), $0.41$ (approximately, the critical density) (green squares), and $1$ (red triangles), where $v_*\rightarrow \infty$ in all three cases. The condensate accommodates the excess mass of amount $L(\rho-\rho_c)$, whereas the mass of amount $L \rho_c$ is distributed according to a $m^{-5/2}$ power law in the bulk.

Now we use the numerically calculated $P(m)$ and $\phi(v)$ [as in Eq.~(\ref{phi})] in Eqs.~\eqref{g_i_general} and \eqref{u_i_general}, to calculate the two observables $g(\rho)$ and $u(\rho)$ as a function of density $\rho$. Then using $u(\rho)$ and $g(\rho)$ in Eq.~(\ref{transportKchip}), we readily obtain the two density-dependent transport coefficients - the bulk diffusion coefficient $D(\rho)$ and the conductivity $\chi(\rho)$.
In Fig. \ref{UgLHMtransport}, we plot $D(\rho)$ (panel (a)) and $\chi(\rho)$ (panel (b)) as a function of density $\rho$ for $v_*=2$ (red dotted lines), 10 (magenta dashed-dotted line) and 100 (green dashed-dotted-dotted line) and for $v_*\rightarrow \infty$ (black dashed line). We then compare the results with that for ZRP (sky-blue solid line) [Eqs.~(\ref{ZRP_diff2}) and (\ref{ZRP_cond2})]. Interestingly, we again observe a nonmonotonic behavior of the bulk-diffusion coefficient $D(\rho)$ with increasing $v_*$. However, the diffusivity remains always finite and never vanishes at any finite density. On the other hand, the conductivity $\chi(\rho)$ monotonically increases with increasing $v_*$ and eventually diverges, in the limit of $v_*\rightarrow \infty$, at the critical density $\rho_c=\sqrt{2}-1$. The observation indeed strongly suggests a direct connection between the conductivity and the cluster formation in the system, thus supporting the scenario of a mobility-driven clustering in the systems.

\subsubsection{Density relaxation and verification of the Einstein relation}
\label{SecProfileUgLHM}

\begin{figure}[h]
\begin{center}
\leavevmode
\includegraphics[width=9cm,angle=0]{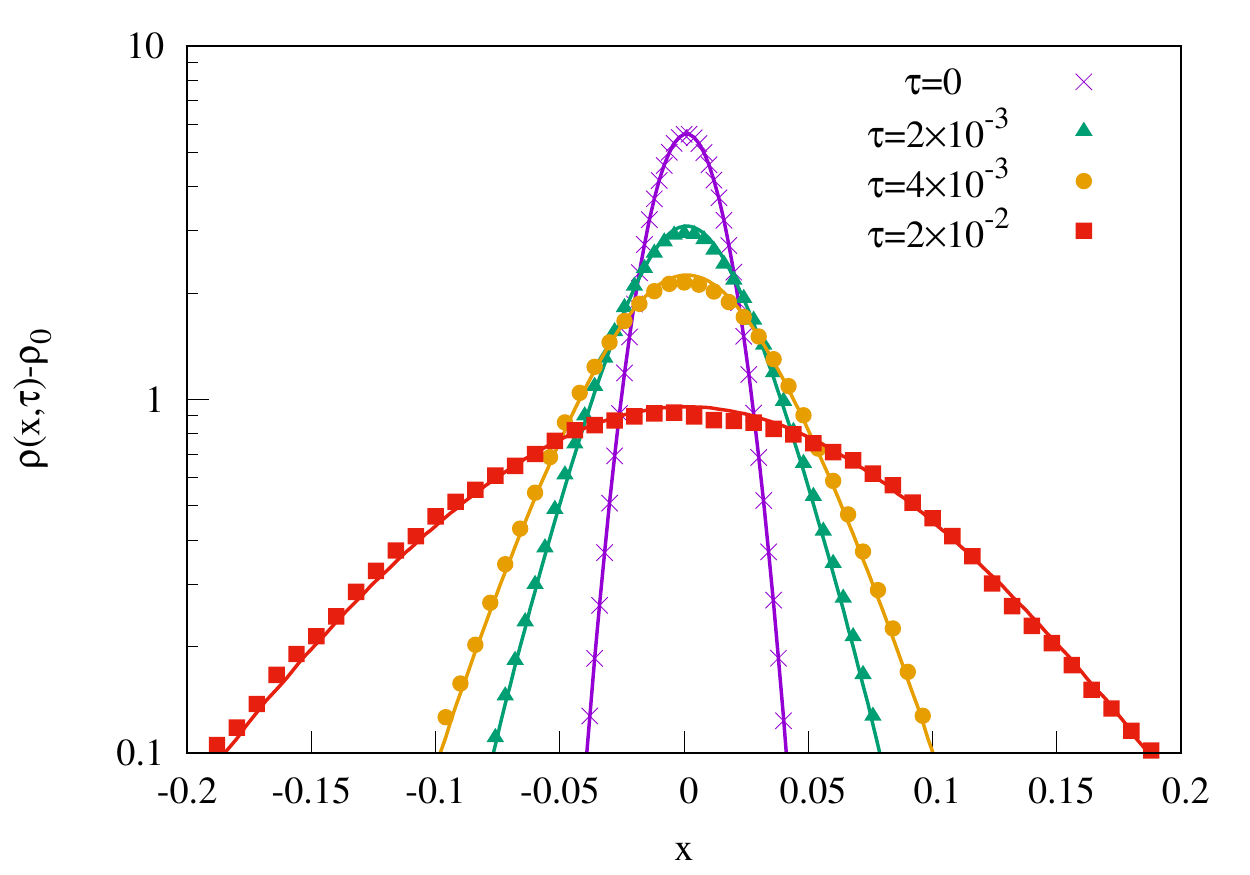}
\caption{ {\it Density relaxation in variant II, $p=q=1/2$:} The relaxation of an initial density perturbation is compared for the final-time density profiles, calculated from hydrodynamic theory and simulations at different times $\tau=0$ (magenta cross), $2\times 10^{-3}$ (green triangle), $4\times 10^{-3}$ (yellow circle), and $2\times 10^{-2}$ (red square) starting from initial condition Eq.~(\ref{initialKchip}) for $v_*=10$, $\rho_0=0.5$, $n_1=0.2$ and $\Delta^2=2\times10^{-4}$. Lines - hydrodynamic theory; points - simulations.}
\label{FigHydroUgLHM}
\end{center}
\end{figure}

We follow the same numerical procedure as in Sec. \ref{SecER_ULHM} to verify the functional dependence of the bulk-diffusion coefficient $D(\rho)$ on density $\rho$ by studying the relaxation of density profiles from an initial density perturbation. For this purpose, we set $\rho_0=0.5$, $n_1=0.2$, and $\Delta^2=2\times10^{-4}$. In Fig. \ref{FigHydroUgLHM}, we compare the density profiles obtained by numerically integrating hydrodynamic time-evolution Eq.~(\ref{hydroKchip5}) and that obtained from microscopic simulations, at various hydrodynamic times $\tau=0$ (magenta cross), $2\times 10^{-3}$ (green triangle), $4\times 10^{-3}$ (yellow circle), and $2\times 10^{-2}$ (red square), starting from initial condition given by Eq.~(\ref{initialKchip}). One can see that the hydrodynamic theory (lines) captures the simulation results (points) quite well.

\begin{figure}[h]
\begin{center}
\leavevmode
\includegraphics[width=9cm,angle=0]{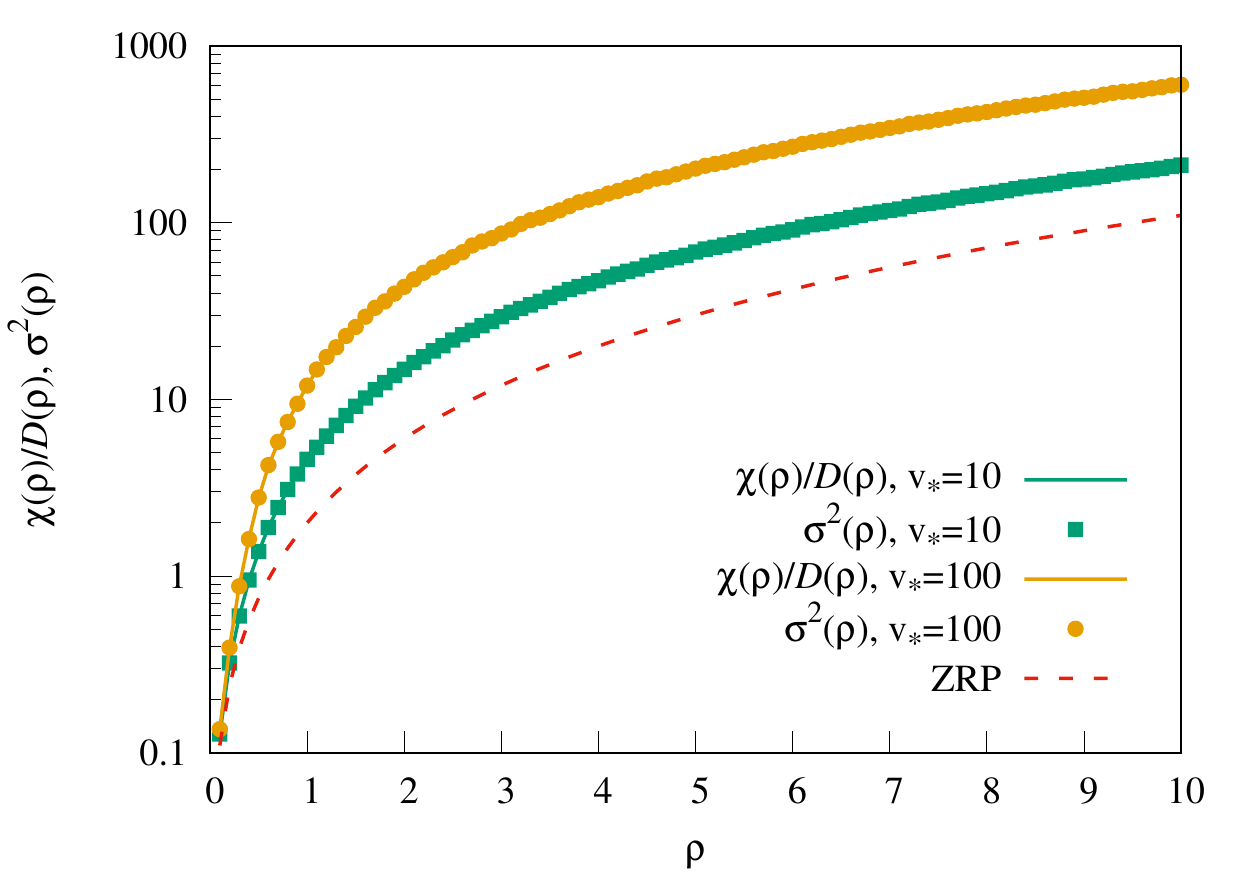}
\caption{ {\it Verification of the Einstein relation in variant II, $p=q=1/2$ }: Scaled subsystem mass fluctuation $\sigma^2(\rho)$ is plotted as a function of density $\rho$ for $v_*=10$ (green square) and 100 (yellow circle). It is compared with the ratio of two transport coefficients $\chi(\rho)$ and $D(\rho)$, calculated numerically using Eq.~(\ref{transportKchip}), for $v_*=10$ (green solid line) and $100$ (yellow solid line). Hydrodynamic theory (lines) and simulations (points) agree quite well, thus demonstrating the existence of the Einstein relation Eq.~(\ref{ER}) in the system. For comparison, the scaled mass fluctuation $\sigma^2(\rho)$ for the ZRP is also plotted (red dashed line), indicating mass fluctuations and the conductivity both grow with  increasing $v_*$.}
\label{UgLHM_ER}
\end{center}
\end{figure}

Finally, we check the validity of the Einstein relation, that connects the scaled mass fluctuation to the ratio of the conductivity and the bulk-diffusion coefficient. In Fig. \ref{UgLHM_ER}, we plot the ratio of the numerically calculated transport coefficients $\chi(\rho)/D(\rho)$ for $v_*=10$ (green solid line) and $100$ (yellow solid line) and compare them with the scaled variance $\sigma^2(\rho)$ of subsystem mass for $v_*=10$ (green square) and 100 (yellow circle) obtained from direct simulations. For comparison, we also plot the scaled mass fluctuation $\sigma^2(\rho)$ for the ZRP (red dashed line). We observe excellent agreement between lines (ratio of transport of transport coefficients) and points (mass fluctuations), thus substantiating the existence of an equilibrium-like Einstein relation in this variant of mass aggregation processes. One should note that the mass fluctuations increase with the increasing ratio of the conductivity to the diffusivity, implying a mobility-driven clustering in the system.

\section{Summary and concluding remarks}
\label{conclusion}

\begin{figure}[h]
\begin{center}
\leavevmode
\includegraphics[width=9cm,angle=0]{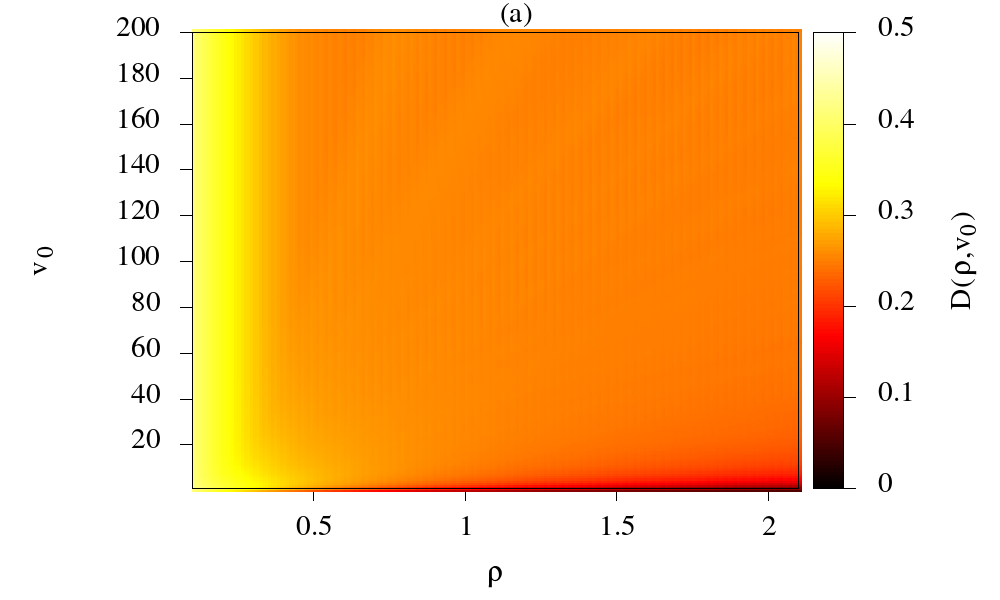}
\includegraphics[width=9cm,angle=0]{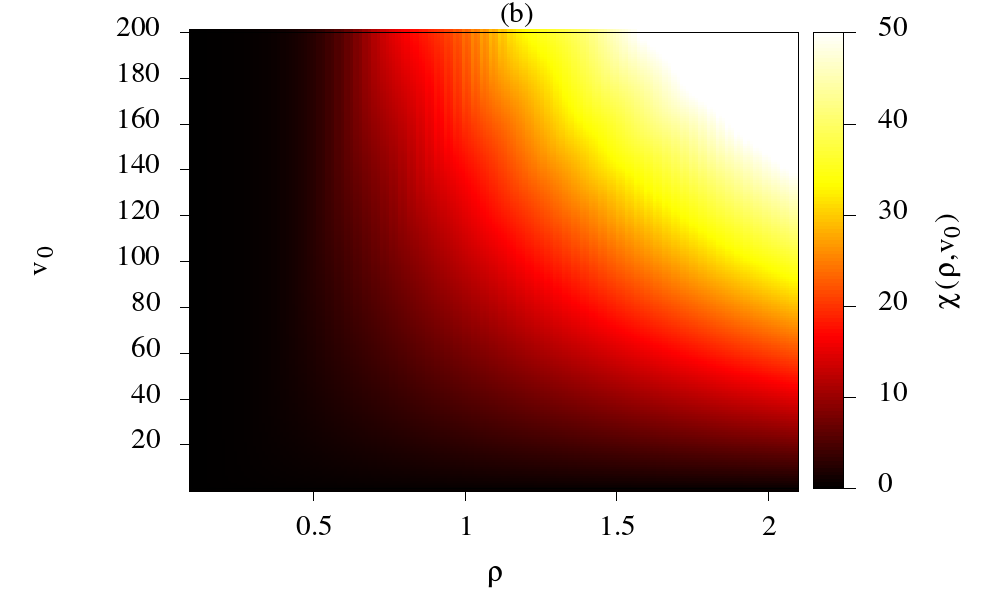}
\caption{ {\it Heat-maps for variant I with $p=q=1/2$}: Two transport coefficients, the bulk-diffusion coefficient $D(\rho, v_0)$ in panel (a) and the conductivity $\chi(\rho, v_0)$ in panel (b), are represented over the parameter space of $\rho$ and $v_0$. }
\label{Heat-map-var1}
\end{center}
\end{figure}

\begin{figure}[h]
\begin{center}
\leavevmode
\includegraphics[width=9cm,angle=0]{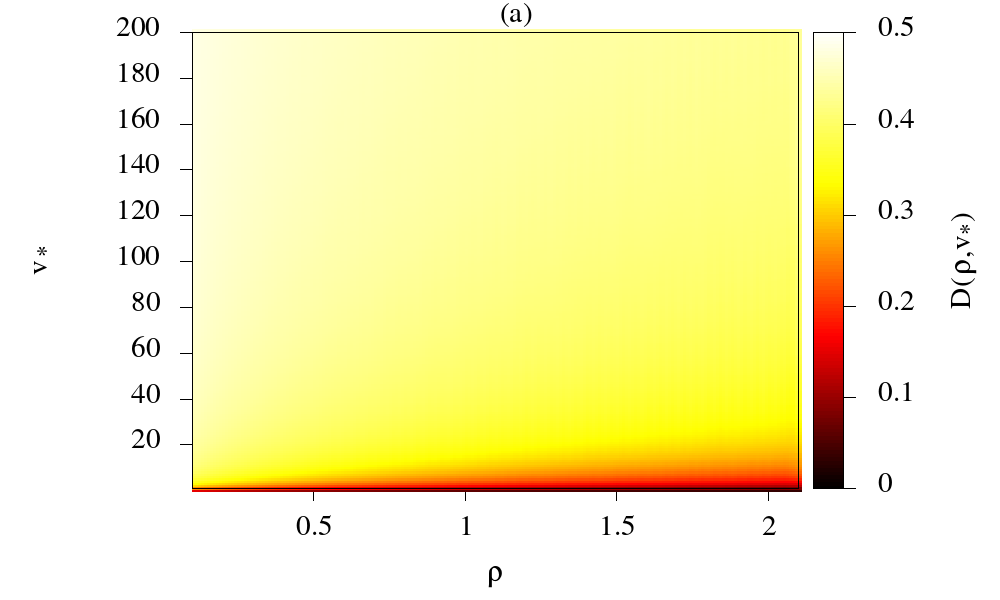}
\includegraphics[width=9cm,angle=0]{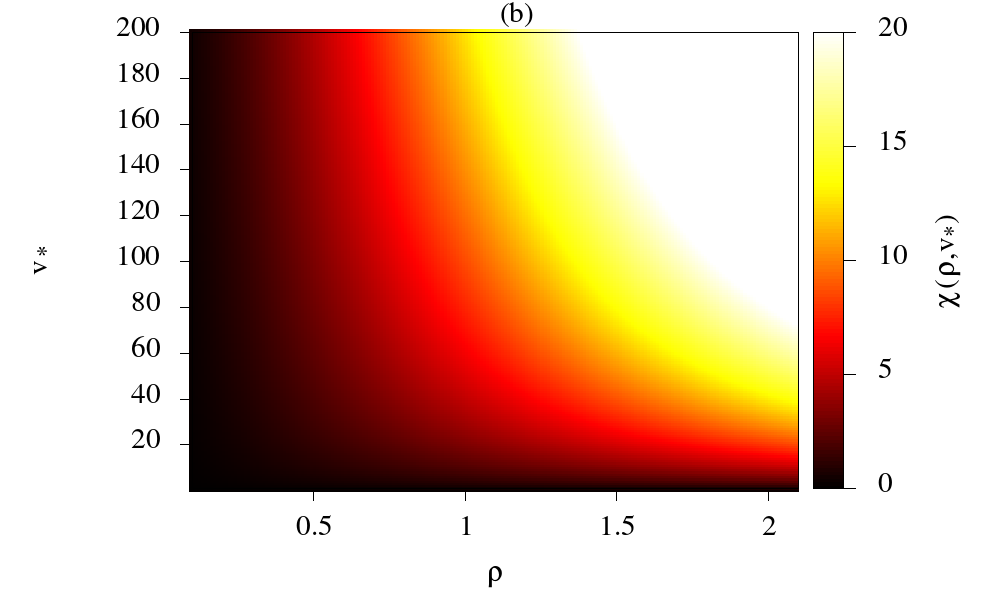}
\caption{ {\it Heat-maps  for variant II with $p=0$, $q=1$:} Two transport coefficients, the bulk-diffusion coefficient $D(\rho, v_*)$ in panel (a) and the conductivity $\chi(\rho, v_*)$ in panel (b), are represented over the parameter space of $\rho$ and $v_*$.} 
\label{Heat-map-var2}
\end{center}
\end{figure}

In this paper, we have studied transport and fluctuation properties of a broad class of one dimensional conserved-mass aggregation processes, which involve fragmentation, diffusion and aggregation of masses \cite{Barma_PRL1998, Barma_JSP2000}. These mass aggregation models, and their variants, have been intensively studied in the past couple of decades, but their hydrodynamic structures are still largely unexplored.
In this scenario, we have calculated two density-dependent transport coefficients, the bulk-diffusion coefficient and the conductivity, which govern the hydrodynamic time-evolution of the density field in the mass aggregation models. We observe that the models have a gradient property, which enables us to identify the local diffusive and drift currents and consequently the two transport coefficients in terms of single-site mass distributions.  In the absence of the knowledge of the steady-state probability weights of microscopic configurations, explicit calculations of the transport coefficients as a function of density are in general difficult. However, in a few special cases, we use a mean-field theory to obtain the steady-state mass distributions, which are then used to calculate the transport coefficients. Indeed, the finite-size scaling analysis (see Appendix \ref{app_3}) suggests that the neighboring spatial correlations vanish in the limit of  large system size, implying that the mean-field expressions of the transport coefficients are exact. We find that the analytic results agree quite well with simulations.

To get a broad overview of our results and to qualitatively understand the role of the various parameters such as fragmentation-cum-aggregation processes and density on the behaviors of the transport coefficients, in Figs. \ref{Heat-map-var1} and \ref{Heat-map-var2} we represent the bulk-diffusion coefficient $D(\rho, v_0)$ [likewise $D(\rho, v_*)$] and the conductivity $\chi(\rho, v_0)$ [likewise $\chi(\rho, v_*)$] as a function of $\rho$ and $v_0$ (likewise $v_*$) through heat-maps for variants I (with  $p=q=1/2$) and II (with $p=0$ and $q=1$); variant II with $p=q=1/2$ has qualitatively similar behavior as that of variant I (not shown in the heat maps). In both the cases, the bulk-diffusion coefficients $D(\rho, v_0)$ and $D(\rho, v_*)$ are bounded. On the other hand, upon increasing the fragmentation size $v_0$ or $v_*$, the conductivity can grow without bound even at moderate density values (see the top-right corner of panel (b) in Figs. \ref{Heat-map-var1} and \ref{Heat-map-var2}).  
In other words, when the fragmentation and aggregation processes dominate (i.e., $v_0$ and $v_*$ are large), the conductivity or, equivalently, the mobility of the particles is greatly enhanced, implying onset of the collective transport, which can lead to a mobility-driven clustering in the systems. In fact, in certain parameter regime, the conductivity, along with the mass fluctuation, even diverges at a critical density, beyond which, there is a macroscopic-size mass condensate,  coexisting with the bulk fluid, which offers essentially zero resistance to the particle flow induced by an external force field. Indeed, analogous to the familiar Bose-Einstein condensation phenomenon, one could think of the system undergoing  a ``superfluid-like'' dynamical transition from a disordered fluid phase having {\it nonzero} resistance to a translation-symmetry broken condensate phase having {\it zero} resistance. Thus we have provided a hydrodynamic characterization of a nonequilibrium condensation transition in terms of a singularity (a simple pole) in the conductivity, which manifests itself through a huge enhancement in the collective particle transport and, for suitable parameter values, induces diverging mass fluctuations in the system.

Note that the mass aggregation models are examples of systems far from equilibrium as they violate, for generic parameter values, the Kolmogorov criterion of microscopic  time-reversibility. Yet, the transport coefficients and the mass fluctuations are found to be connected by an equilibrium-like Einstein relation - a corner-stone for formulating a fluctuating hydrodynamics framework for equilibrium systems. Indeed, hydrodynamics of these nonequilibrium aggregation processes can lead to the formulation of macroscopic fluctuation theory \cite{Bertini_JSP2002, Bertini_RMP2015}, which will help in understanding fluctuation properties  of these systems under various driving conditions.

In the previous studies of mass aggregation processes, mainly the static properties of the condensation transition and the related mass clustering have been studied  \cite{Barma_PRL1998, Puri}.  In this work, we have studied the time-dependent properties of these mass-aggregating systems in terms of relaxations of density profiles from given initial conditions. Indeed, by calculating the transport coefficients, which govern the time evolution of the density profiles, we have demonstrated that the dynamical origin of the condensation transition lies in the diverging conductivity, not the vanishing diffusivity. In other words, unlike the dynamical slowing-down, which is usually observed at the critical point for equilibrium systems, the nonequilibrium phase transition studied here is driven by a huge enhancement of the particle mobility. We believe that this particular mechanism of mobility-driven clustering could be the signature of not only the aggregation related clustering phenomena, but also the clustering observed in various active-matter systems \cite{Goldstein_PRL2004, Lopez_prl2015, Golestanian_pre2019, Tanmoy_longhop, Marconi_PRL2020}. From an overall perspective, the mechanism could provide an exciting avenue for characterizing phase transitions in a broad class of out-of-equilibrium systems.

Generalization of the results to higher dimensions should be straightforward.
There are a few open issues though. For simplicity, in this work we have considered only the mass-independent rates for fragmentation, diffusion and aggregation. However, one can in principle generalize the models where fragmentation and diffusion rates depend on the masses at the departure sites and the aggregation rates, considered in the literature through a mass-dependent kernel \cite{Ziff_PRL1982}, depend on the masses at both the departure and the destination sites. In the first case, the system would still possess a gradient structure and the transport coefficients can be formally expressed in terms of the single-site mass distributions. However, obtaining the analytic expressions of the transport coefficients may be difficult as the steady-state probabilities of the microscopic configurations are not known. In the latter case, the systems have a non-gradient structure and calculating the transport coefficients remains a challenge.

\section{Acknowledgment}

We thank R. Rajesh for discussion. P.P. acknowledges the Science and Engineering
Research Board (SERB), India, under Grant No. MTR/2019/000386, for financial support.  T.C. acknowledges a research fellowship [Grant No. 09/575 (0124)/2019-EMR-I] from the Council of Scientific and Industrial Research (CSIR), India. T.C., S.C. and A.D. acknowledge the hospitality at the International Centre for Theoretical Sciences (ICTS), Bengaluru during the Bangalore School on Statistical Physics - X  (Code: ICTS/bssp2019/06), where part of the work was done.

\section{Appendix}

\subsection{ Time evolution of local density}
\label{app_1}

 Here we provide calculation details of deriving the time evolution of density at site $i$ in the presence of a biasing force $F$ along $+x$ direction. Introducing the biased hopping rates as shown in Eq.~(\ref{bias4}), the infinitesimal-time evolution of mass $m_i(t)$ can be written as
\begin{eqnarray} \nonumber
m_i(t+dt) = ~~~~~~~~~~~~~~~~~~~~~~~~~~~~~~~~~~~~~~~~~~~~~~~~~~~~~ \\ 
\left\{
\begin{array}{ll}
m_i(t) - 1          & {\rm prob.}~ pW_{i,i+1}^{F}\hat{a}_i dt/2 \\
m_i(t) - 1          & {\rm prob.}~ pW_{i,i-1}^{F}\hat{a}_i dt/2 \\
m_i(t) + 1          & {\rm prob.}~ pW_{i-1,i}^{F} \hat{a}_{i-1} dt/2 \\
m_i(t) + 1          & {\rm prob.}~ pW_{i+1,i}^{F} \hat{a}_{i+1} dt/2 \\
m_i(t) - v          & {\rm prob.}~ qD_{i,i+1}^{F,v}\hat{a}_i^v \phi(v) dt/2 \\
m_i(t) - v          & {\rm prob.}~ qD_{i,i-1}^{F,v}\hat{a}_i^v \phi(v) dt/2 \\
0                   & {\rm prob.}~ qD_{i,i+1}^{F}\hat{a}_i(1-\hat{a}_i^v) \phi(v) dt/2 \\
0                   & {\rm prob.}~ qD_{i,i-1}^{F}\hat{a}_i(1-\hat{a}_i^v) \phi(v) dt/2 \\
m_i(t) + m_{i-1}(t) & {\rm prob.}~ qD_{i-1,i}^{F}\hat{a}_{i-1} (1-\hat{a}_{i-1}^v) \phi(v) {dt}/{2} \\
m_i(t) + v          & {\rm prob.}~ qD_{i-1,i}^{F,v} \hat{a}_{i-1}^v \phi(v) dt/2 \\
m_i(t) + m_{i+1}(t) & {\rm prob.}~ qD_{i+1,i}^{F}\hat{a}_{i+1} (1-\hat{a}_{i+1}^v) \phi(v) {dt}/{2} \\
m_i(t) + v          & {\rm prob.}~ qD_{i+1,i}^{F,v} \hat{a}_{i+1}^v \phi(v) dt/2 \\
m_i(t)              & {\rm prob.}~ 1-\Sigma_F dt
\end{array}
\right.
\label{General_Update_modify}
\end{eqnarray}
with
\begin{align*}
&\Sigma_F = \frac{p}{2} \left[\hat{a}_i \left(W_{i,i+1}^{F}+W_{i,i-1}^{F}\right) + W_{i-1,i}^{F} \hat{a}_{i-1} + W_{i+1,i}^{F} \hat{a}_{i+1} \right] \nonumber \\
&+ \frac{q}{2} \sum_{v=0}^\infty \phi(v) \left[ \hat{a}_i^v \left(D_{i,i+1}^{F,v} + D_{i,i-1}^{F,v}\right) + \hat{a}_i(1-\hat{a}_i^v) \left(D_{i,i+1}^{F} \right. \right. \nonumber \\
&+ \left. D_{i,i-1}^{F} \right) + D_{i-1,i}^{F}\hat{a}_{i-1} (1-\hat{a}_{i-1}^v) + D_{i-1,i}^{F,v} \hat{a}_{i-1}^v  \nonumber \\ 
&+ D_{i+1,i}^{F}\hat{a}_{i+1} (1-\hat{a}_{i+1}^v) + \left. D_{i+1,i}^{F,v} \hat{a}_{i+1}^v \right]. \nonumber
\end{align*}
The quantities $W_{i,i+1}^{F}$, $D_{i,i+1}^{F}$ and $D_{i,i+1}^{F,v}$ denote the modified (biased) mass transfer rates from site $i$ to $i+1$ in the cases of chipping of a single-unit mass, fragmentation of whole mass $m_i$ and fragmentation of mass $v$, respectively, and are written below after using the linearized form as in Eq.~(\ref{bias4}),
\begin{align}
W_{i,j}^{F} &= 1 + \frac{1}{2} F(j-i), \\
D_{i,j}^{F,v} &= 1 + \frac{1}{2}vF(j-i), \\
D_{i,j}^{F} &= 1 + \frac{1}{2}m_i F(j-i). \\ \nonumber
\end{align}
From the update rules as given in Eq.~(\ref{General_Update_modify}), one can determine the time evolution equation for average mass $\left\langle m_i (t)\right\rangle = \rho_i(t)$ at site $i$,  
\begin{widetext}
\begin{align*}
& \frac{d\left\langle m_i \right\rangle}{dt} = \frac{p}{2} \left[\left(W_{i,i+1}^{F}+W_{i,i-1}^{F}\right) \left\langle (m_i(t) - 1) \hat{a}_i  \right\rangle +W_{i-1,i}^{F}\left\langle (m_i(t) + 1)  \hat{a}_{i-1} \right\rangle + W_{i+1,i}^{F} \left\langle (m_i(t) + 1)  \hat{a}_{i+1} \right\rangle \right. \\
& \left. - \left(W_{i,i+1}^{F}+W_{i,i-1}^{F}\right)\left\langle m_i(t) \hat{a}_i \right\rangle - W_{i-1,i}^{F} \left\langle m_i(t) \hat{a}_{i-1} \right\rangle  - W_{i+1,i}^{F} \left\langle m_i(t) \hat{a}_{i+1} \right\rangle \right] + \frac{q}{2} \sum_{v=0}^\infty \phi(v) \left[  \left(D_{i,i+1}^{F,v} + D_{i,i-1}^{F,v}\right)\left\langle (m_i(t) - v)\hat{a}_i^v  \right\rangle \right. \\
& \left. + \left\langle (m_i(t) + m_{i-1}(t)) D_{i-1,i}^{F}\hat{a}_{i-1} (1-\hat{a}_{i-1}^v) \right\rangle + D_{i-1,i}^{F,v} \left\langle (m_i(t) + v) \hat{a}_{i-1}^v \right\rangle + \left\langle (m_i(t) + m_{i+1}(t)) D_{i+1,i}^{F}\hat{a}_{i+1} (1-\hat{a}_{i+1}^v) \right\rangle  \right. \\ \\
& \left. + D_{i+1,i}^{F,v} \left\langle (m_i(t) + v) \hat{a}_{i+1}^v \right\rangle - \left\langle m_i(t)  \left\{ \left(D_{i,i+1}^{F} + D_{i,i-1}^{F} \right) \hat{a}_i (1-\hat{a}_i^v)  + D_{i-1,i}^{F}\hat{a}_{i-1} (1-\hat{a}_{i-1}^v) +  D_{i+1,i}^{F}\hat{a}_{i+1} (1-\hat{a}_{i+1}^v) \right\} \right\rangle  \right. \\ \\
& \left. - \left(D_{i,i+1}^{F,v} + D_{i,i-1}^{F,v}\right) \left\langle m_i(t) \hat{a}_i^v \right\rangle - D_{i-1,i}^{F,v}\left\langle m_i(t) \hat{a}_{i-1}^v \right\rangle - D_{i+1,i}^{F,v}\left\langle m_i(t) \hat{a}_{i+1}^v \right\rangle \right]. 
\end{align*}
\end{widetext}

Now by using the identity $\langle m_i \hat{a}_i \rangle = \langle m_i (1-\delta_{m_i,0}) \rangle = \langle m_i \rangle = \rho_i$, substituting the modified mass transfer rates $W_{i,i+1}^{F}$, $D_{i,i+1}^{F}$ and $D_{i,i+1}^{F,v}$ in the above equation and after some algebraic manipulations, we obtain
\begin{widetext}
\begin{eqnarray} \nonumber
\frac{\partial \rho_i}{\partial t} &=& \frac{1}{2} \left[ p \left\langle (\hat{a}_{i-1} + \hat{a}_{i+1} - 2\hat{a}_i) \right\rangle + q (\rho_{i-1} + \rho_{i+1} - 2\rho_i) \right] \\ \nonumber
&& + \frac{q}{2} \sum_{v=0}^\infty \phi(v) \left[ \left\langle v (\hat{a}_{i+1}^v + \hat{a}_{i-1}^v - 2\hat{a}_i^v) \right\rangle - \left(\left\langle m_{i+1}\hat{a}_{i+1}\hat{a}_{i+1}^v \right\rangle + \left\langle m_{i-1}\hat{a}_{i-1}\hat{a}_{i-1}^v \right\rangle - 2\left\langle m_i \hat{a}_i  \hat{a}_i^v \right\rangle \right)\right] 
\\ \nonumber
&& + \frac{F\delta x}{4} \left[ p\left\langle \hat{a}_{i-1} \right\rangle - q\sum_{v=0}^\infty \phi(v) \left\langle m_{i-1}^2 \hat{a}_{i-1} \hat{a}_{i-1}^v \right\rangle + q\left\langle m_{i-1}^2 \right\rangle + q\sum_{v=0}^\infty \phi(v) \left\langle v^2 \hat{a}_{i-1}^v \right\rangle \right] \\
&& - \frac{F\delta x}{4} \left[ p\left\langle \hat{a}_{i+1} \right\rangle - q\sum_{v=0}^\infty \phi(v) \left\langle m_{i+1}^2 \hat{a}_{i+1}\hat{a}_{i+1}^v \right\rangle + q\left\langle m_{i+1}^2 \right\rangle + q\sum_{v=0}^\infty \phi(v) \left\langle v^2 \hat{a}_{i+1}^v \right\rangle \right].
\label{density-evoln}
\end{eqnarray}

\end{widetext}

\subsection{Violation of Detailed Balance}
\label{app_2}

{\it Variant I, $v_0=2$.} For illustration, we first consider the case of mass aggregation model -  variant I with $v_0=2$ and $p=q=1/2$. Detailed balance (DB) is violated if there exists a pair of configurations $C_1$ and $C_2$ such that the probability current,
\begin{eqnarray}\label{DB_v2}
\Delta J= W_{C_1, C_2} P(C_1) - W_{C_2, C_1} P(C_2) \neq 0 ,
\end{eqnarray}
is non-zero.
Here we denote $W_{C_1, C_2}$ as the transition rate from configuration $C_1$ to $C_2$ and $P(C_1)$ and $P(C_2)$ are the steady-state probabilities of the two configurations $C_1$ and $C_2$, respectively. To show the violation of DB, we consider two nearest neighbor sites $i$ and $i+1$ and configurations $C_{1}\equiv (m_1, m_2, \dots , m_{i}=m, m_{i+1}=m', \dots)$ and $C_{2}\equiv (m_1, m_2, \dots , m_{i}=m-1, m_{i+1}=m'+1, \dots)$ with $m>0$. In this case, the transition from $C_{1}$ to $C_{2}$ (and the reverse one) is solely contributed by  unit-mass transfer across the bond ($i, i+1$), where the transition rates are given by
  \begin{eqnarray}
  W_{C_1,C_2}=\frac{p}{2}+ \frac{q}{2} \delta_{m,1}, \nonumber \\
  W_{C_2,C_1}=\frac{p}{2}+ \frac{q}{2} \delta_{m',0}.
  \end{eqnarray}
 As the vanishing neighboring correlations (see Fig. \ref{Fig:Scaled_correlation_var1}) suggest a statistical independence of neighboring sites, the steady-state joint mass distribution can be written in a product form. That is, the probability of configuration $C_{1}  \equiv \{ m_1, \dots, m, m', \dots m_L \}$ is given by
  \begin{eqnarray}
  P(C_1)&=& P(m_1) \dots P(m_{i}=m) P(m_{i+1}=m')\dots P(m_L) \nonumber \\
        &=&  \kappa  P(m) P(m'),
        \end{eqnarray}
 where we denote $\kappa = \prod_{\substack{s=1 \\ s \neq i, i+1 }}^{L} P(m_s)$. Similarly, for configuration $C_2$, the configuration probability can be written as
  \begin{eqnarray}
       P(C_2)=\kappa  P(m-1) P(m'+1).
 \end{eqnarray}
 Considering $p=q=1/2$ and using the single-site mass distribution as in Eq.~(\ref{distn_v2}), the lhs of Eq.~(\ref{DB_v2}) can be written as
\begin{eqnarray}
\Delta J =  \kappa\frac{P_1 P_0}{4} \left[\delta_{m,1} \left(\frac{P_1}{P_0}\right)^{m'} F_{m'+1} - \delta_{m',0} \left(\frac{P_1}{P_0}\right)^{m-1} F_{m} \right] +  \nonumber \\  
 \kappa\frac{P_0^{2}}{4} \left(\frac{P_1}{P_0}\right)^{m+m'} \left[ F_{m+1}F_{m'+1} - F_{m}F_{m'+2} \right], \nonumber \\
\end{eqnarray}
which is in general non-zero. This can be simply seen by considering a  case where $m=1$. Then the above equation is simplified to
\begin{eqnarray}\label{DB_v2_calc}
\Delta J =\kappa\frac{P_0^{2}}{4} \left(\frac{P_1}{P_0}\right)^{1+m'} \left[ F_{m'+1}-F_{m'} \right] - \kappa\frac{P_1 P_0}{4} \delta_{m',0}. \nonumber \\
\end{eqnarray}
 It is easy to check that the rhs of Eq.~(\ref{DB_v2_calc}) vanishes for $m'= 0$, $1$ and gives a non-zero contribution for any other values of $m'$. Therefore, the forward and backward mass-transfer events with the above mentioned configurations $C_1$ and $C_2$ with $m=1$ and $m'>1$ lead to the violation of DB.

{\it Variant I, $v_0=\infty$.} Proving violation of DB is simple in this case. Consider an aggregation event of two neighboring masses $m$ and $m'$, which become a single mass of amount $m+m'$. However, the reverse process is not possible, implying violation of Kolmogorov criterion and therefore violation of DB. In a similar way, one could also show violation of DB for any other $v_0$.

\subsection{Finite-size scaling of correlation functions}
\label{app_3}

\begin{figure}[H]
\begin{center}
\leavevmode
{\includegraphics[width=9cm,angle=0]{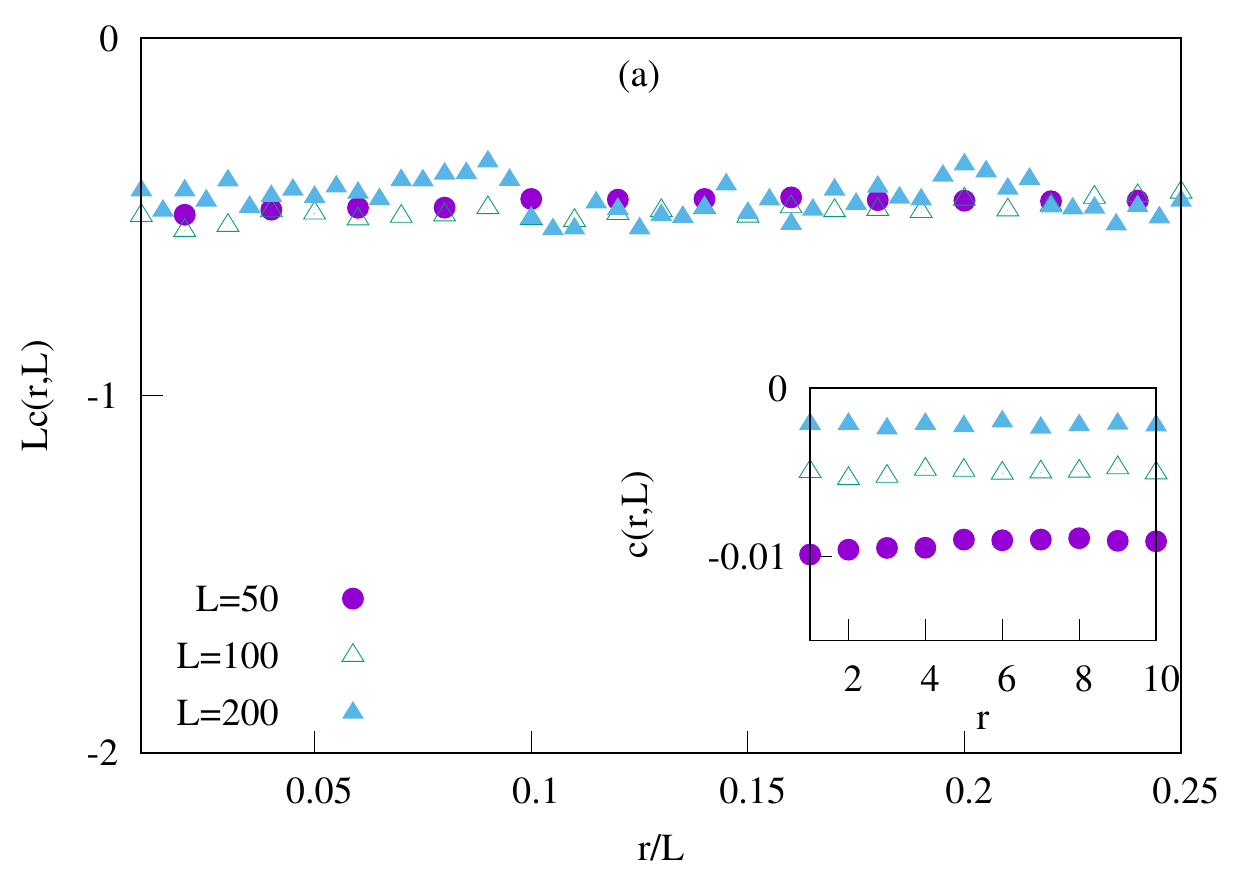}} \hfill
{\includegraphics[width=9cm,angle=0]{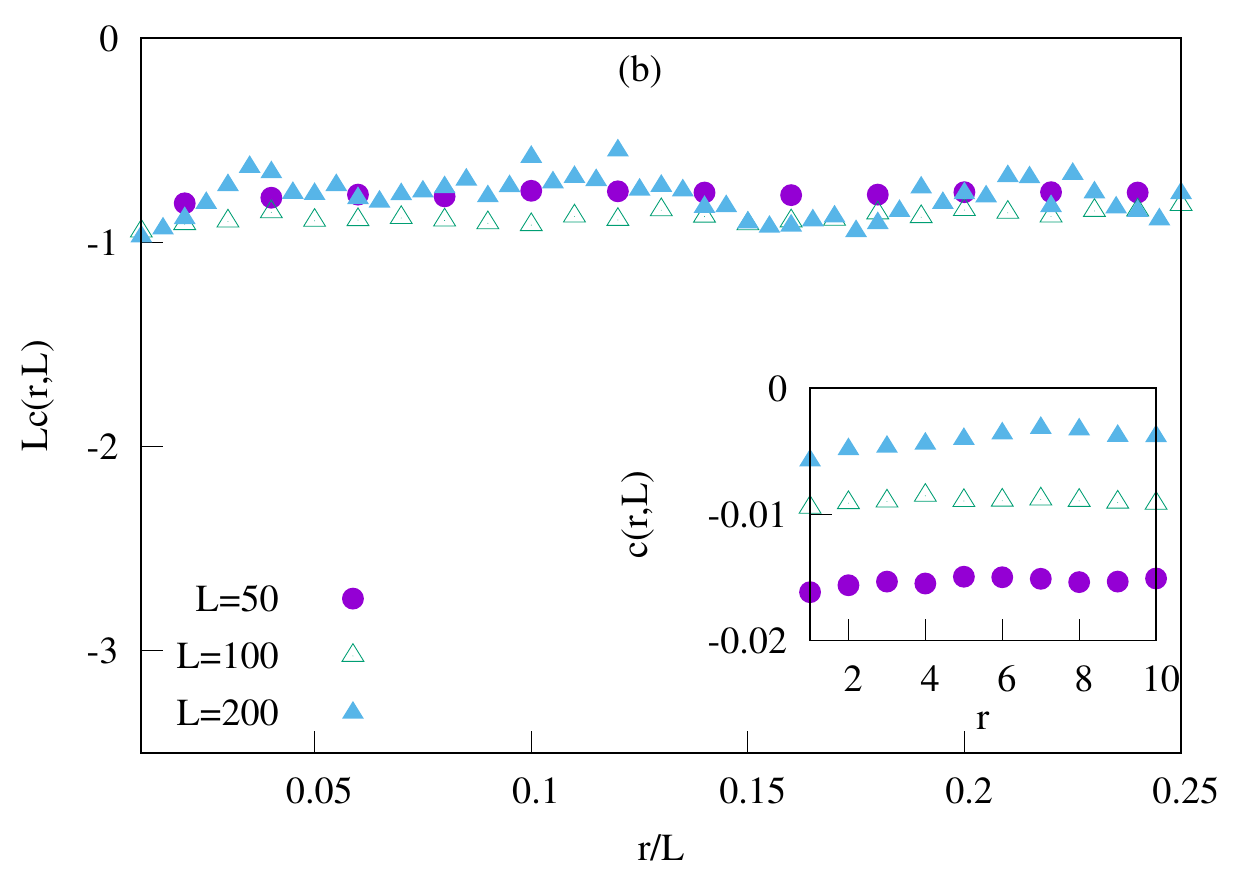}}
\caption{ {\it Variant I, $p=q=1/2$.}  Scaled two-point spatial correlation function $L c(r,L)$ is plotted as a function of scaled distance $r/L$ for density $\rho=0.3$ and for $v_0=2$ [panel (a)] and $v_0=\infty$ [panel (b)]. The above simulation is performed for $L= 50$ (filled circle), $100$ (open triangle) and $200$ (filled triangle) \textit{Inset:} We plot (unscaled) correlation function $c(r,L)$ as a function of distance $r$ for above mentioned system sizes. 
We find $c(r,L) \sim {\cal O}(1/L)$ approaches zero with increasing $L$. }
\label{Fig:Scaled_correlation_var1}
\end{center}
\end{figure}

\begin{figure}[h]
\centering
{\includegraphics[width=1 \linewidth]{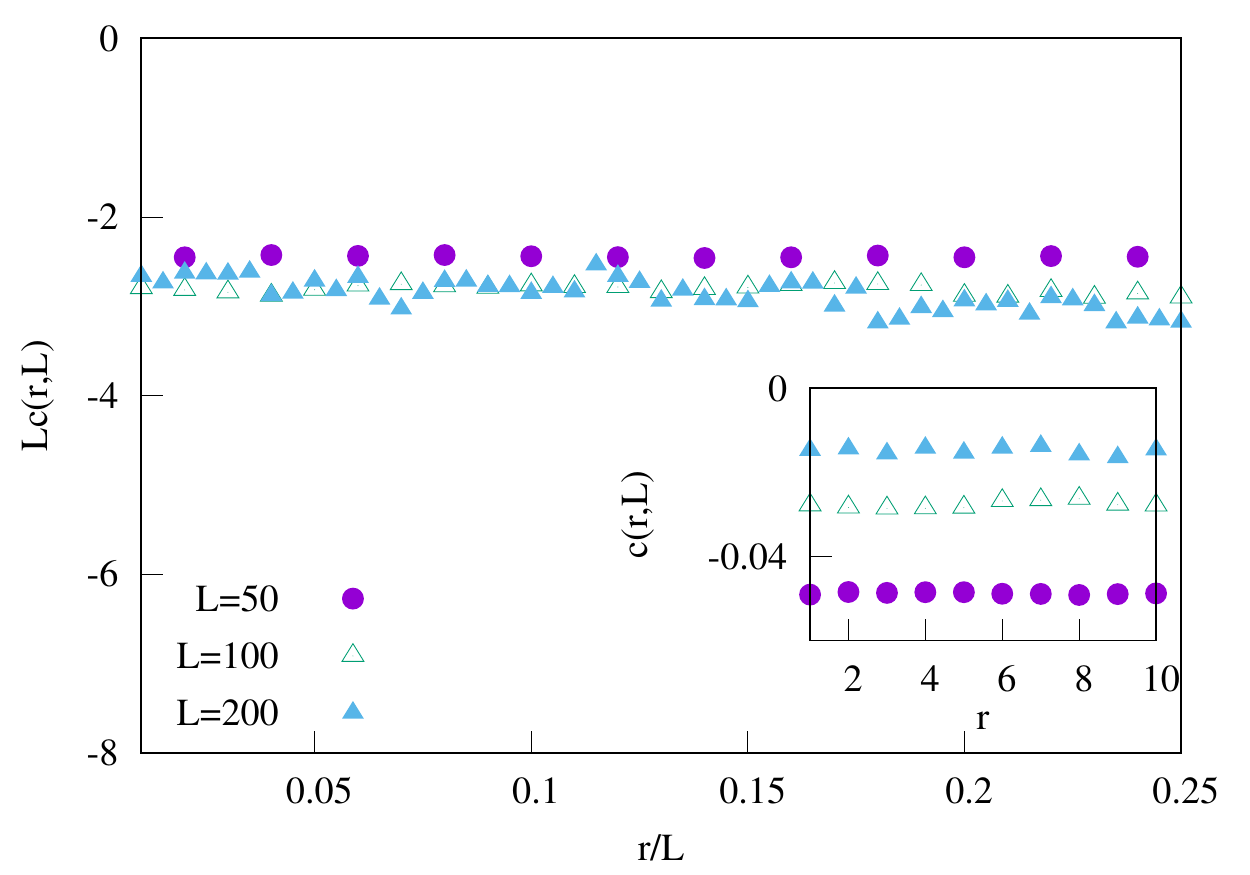}}\hfill
\caption{ {\it Variant II.}  Scaled two-point spatial correlation function, $Lc(r,L)$ is plotted as a function of normalized spatial distance $r/L$ for density $\rho=0.3$, $p = 0$, $q=1$ and $v_*=100$. The above simulation is performed for $L=50$ (filled circle), $100$ (open triangle) and $200$ (filled triangle) and the scaling collapse captures the $1/L$ dependence of $c(r,L)$ for large $L$. \textit{Inset:} We plot (unscaled) correlation function $c(r,L)$ as a function of the distance $r$ for above mentioned system sizes. We find $c(r,L)$ approaching towards zero with increasing $L$.}
\label{Fig:Scaled_correlation_var2}
\end{figure}

Here we calculate in simulations the two-point spatial correlation function $c(r, L) = \left\langle m_i m_{i+r} \right\rangle - \rho ^2$ where $L$ is the system size and perform a finite-size scaling analysis. In insets of Figs. \ref{Fig:Scaled_correlation_var1} and \ref{Fig:Scaled_correlation_var2}, we plot, for global density $\rho = 0.3$, spatial correlations $c(r)$ as a function of distance $r$ (where $r \ne 0$) for variant I ($v_0=1$ and $\infty$) and variant II ($v_*=100$). Clearly, for large system size $L \gg 1$, the correlation function $c(r, L)$ is vanishingly small, i.e., $c(r,L) \sim {\cal O}(1/L)$, for all neighboring points with $r \ge 1$;
In Fig. \ref{Fig:Scaled_correlation_var1}, we plot the scaled two-point spatial correlation function $L c(r, L)$ as a function of scaled position $r/L$ for both the variants studied in the paper. The reasonably good scaling collapse suggests that the correlation functions have a scaling form $c(r, L)\simeq (1/L) f(r/L)$ where $f(x)$ is a bounded function of $x$, implying spatial correlations vanish in the thermodynamic limit.

\bibliography{mass_transport1}

\end{document}